\documentclass[twocolumn,amssymb,amsmath,
nofootinbib,tightenlines,showpacs,floatfix,
superscriptaddress,aps,prb]{revtex4-1}
\usepackage{amsfonts}
\usepackage{txfonts}
\usepackage{amsbsy}
\usepackage{bm}
\usepackage[colorlinks=true,linkcolor=blue,filecolor=blue,menucolor=yellow,urlcolor=blue,citecolor=blue,anchorcolor=blue]{hyperref}
\usepackage{graphicx}
\usepackage{times} 
\usepackage{dcolumn}

\newcommand{\bsigma}{\mbox{\boldmath$\sigma$}}

\begin{document}

\title{Spin Transport in Half-Metallic Ferromagnet-Superconductor Junctions}
\author{Chien-Te Wu}
\email{chientewu@nctu.edu.tw}
\affiliation{Department of Electrophysics, National Chiao Tung University, Hsinchu 30010, Taiwan, Republic of China}
\affiliation{Physics Division, National Center for Theoretical Sciences, Hsinchu 30010, Taiwan, Republic of China}
\author{Klaus Halterman }
\email{klaus.halterman@navy.mil}
\affiliation{Michelson Lab, Physics
Division, Naval Air Warfare Center, China Lake, California 93555}
\date{\today}

%%%%%%%%%%%%%%%%%%% abstract and OCIS codes %%%%%%%%%%%%%%%%

\begin{abstract} 
We investigate the charge and spin transport in half-metallic ferromagnet ($F$) and
superconductor ($S$)  nanojunctions. We utilize a self-consistent microscopic method that
can accommodate the broad range of energy scales present, and ensures
proximity effects that account for the interactions at the interfaces are accurately determined.
Two experimentally relevant half-metallic junction 
types are considered: The first is a $F_1 F_2 S$ structure, where
a half-metallic ferromagnet $F_1$ adjoins a weaker conventional ferromagnet $F_2$.
The current is injected  through the $F_1$ layer by means of an applied bias voltage.
The second configuration involves a $S F_1 F_2 F_3 S$ Josephson junction 
whereby a
phase difference $\Delta\varphi$ between the two superconducting electrodes generates the 
supercurrent flow. In this case, the central half-metallic $F_2$ layer is 
surrounded by two weak ferromagnets $F_1$ and
$F_3$.  By placing a ferromagnet with a weak exchange field adjacent to an $S$ layer,
we are able to optimize the conversion process in which opposite-spin triplet pairs are
converted into equal-spin triplet pairs that propagate deep into the half-metallic regions in both
junction types.
For the tunnel junctions, we study the bias-induced local magnetization, spin currents, and spin
transfer torques for various orientations of the relative magnetization angle $\theta$ in the $F$ layers.
We  find that the bias-induced equal-spin triplet pairs are maximized
in the half-metal for $\theta\approx90^\circ$ and
as part of the conversion process, are anticorrelated with the opposite-spin pairs.
We show that the charge current density is maximized, corresponding
to the occurrence of a large amplitude of equal-spin triplet pairs, when the exchange interaction 
of the weak ferromagnet is about $0.1E_F.$
For the half-metallic Josephson junctions we often find that the spin current flowing in the half-metal
is equivalent to the charge supercurrent flowing throughout the junction. This is indicative
that  the current consists of spin-polarized triplet pairs. The conversion process of 
the opposite-spin triplet pairs to the equal-spin triplet pairs in 
the weaker magnets is clearly demonstrated.
This is exemplified by the fact that 
the supercurrent in
the half metal was found to be relatively insensitive to its thickness.
\end{abstract}
\maketitle

\section{Introduction}
\label{intro}
Superconductor ($S$) and ferromagnet ($F$) hybrids have opened up many new possibilities 
for further advancements in spintronics
devices whose purpose is to manipulate the flow of charge and spin currents~\cite{linder}. 
Central to their functionality is experimental control of the spin degree of freedom
while enjoying the dissipationless nature of supercurrent. 
This control is typically afforded through magnetization rotations of 
one of the free ferromagnetic layers, achieved via weak in-plane external magnetic fields, 
or by the spin transfer torque (STT) effect. %ct 
The most commonly studied transport structures 
based on superconductors and ferromagnets
are equilibrium Josephson junctions 
%with ferromagnetic elements, 
or voltage biased superconducting  tunnel junctions.
%with ferromagnetic elements, 
In any case, the underlying junction architecture often involves
spin and charge transport through a  spin-valve configuration.
A basic superconducting spin-valve consists of two or more ferromagnets 
adjacent to a superconductor\cite{oh}, where rotation of one of the $F$ layer 
magnetizations modifies the induced oscillatory
singlet pairing in the ferromagnets. 
If the $F$ layers are 
half-metallic,
these oscillations  rapidly dampen out
due to their incompatible nature.
If however the ferromagnetic regions have  
non-collinear magnetizations, 
as will be discussed shortly,
triplet pairs\cite{berger}
with parallel projection of spin can be created that 
extend deep within the half-metal. 
These  spin-polarized triplet pairs are thus of great interest,
and their 
signatures have been experimentally observed 
in the superconducting critical temperature of half-metallic 
spin valves\cite{robby}
when rotating one of the $F$ layer magnetizations.
Transport measurements in a half-metallic Josephson junction~\cite{keizer} demonstrated 
 a supercurrent through the half-metal
$\rm CrO_2$, also indicating the current is carried by  
equal-spin Cooper pairs since singlet pairs are blocked by the half metal. 
Because control of the transport of dissipationless spin-currents is
a major objective of low-temperature spintronics devices,
superconducting junctions that merge half-metallic ferromagnets and superconductors 
are increasingly being recognized as valuable platforms to study these  two competing 
orders. 

Spin 
currents
can flow within
superconducting  
junctions  
with 
two or more $F$ layers
due to the ferromagnetic exchange 
interactions. 
%and 
They 
can also flow with
the help of induced equal-spin 
triplet pairing correlations, where the
Cooper pairs have a net spin of 
$m = \pm1$ on the spin quantization
axis. 
The generation of these long-range 
triplet correlations in superconducting
heterostructures with magnetic inhomogeneities 
has been well studied theoretically  and experimentally. 
By introducing  magnetic 
inhomogeneity, e.g., inclusion
of multiple magnets
with misaligned exchange
fields, the Hamiltonian no 
longer commutes 
with the total
spin operator 
and equal-spin triplet correlations can be
induced.
Due to the imbalance between majority and
minority spins in a ferromagnet, conventional singlet %superconducting
pairing correlations decay over short distances within  
the magnetic region. However, 
Cooper pairs with electrons
that carry the same spin ($m=\pm 1$) are 
not subject to the paramagnetic pair breaking
and can in principle propagate for large distances inside
the ferromagnet, limited only 
by coherence breaking processes.
Such long-range $m=\pm 1$ triplet correlations 
thus play
an important role in 
Josephson and tunneling junctions containing 
ferromagnets with noncollinear magnetizations.
%Differing degrees of magnetic
%inhomogeneity can be realized 
% by tuning the magnetization
%orientation in each of the magnetic layers.
%This creates an effective triplet conversion
%process and a suitable  platform in which to 
%study  spin-polarized
%currents in superconducting systems.
%

%[Half-metallic basic idea, 
%theory, include triplet discussion here]
While there has been extensive work 
towards isolating and detecting the 
triplet pairing state,
it can be difficult
to disentangle the equal-spin triplet
and opposite-spin singlet  and triplet correlations.
It is therefore of interest to 
investigate heterostructures
that restrict the formation of opposite spin pairs while
retaining the desired equal-spin triplet correlations. 
The pinpointing
of triplet effects can be exploited with the use 
of highly
polarized materials like half metallic 
ferromagnets,
where only a single spin channel is present at the 
Fermi
level. The ordinary singlet 
pairs and opposite-spin 
triplet pairs are consequently  suppressed,
as the magnet behaves essentially 
as an insulator
for the opposite spin band. 
Half-metallic ferromagnets
are thus finding increasing use in
superconducting spin valves~\cite{ah18}.
Several half metallic
materials 
 are considered in connection with 
 superconducting
hybrids\cite{kriv} and spintronic applications. 
These include
the manganese
perovskite ${\rm La_{2/3} Ca_{1/3}Mn O_3}$, 
as well as the 
Heusler compounds such as
$\rm Cu_2MnAl$,
%$\rm Co_2 Fe Si$ and $\rm Co_2 Mn Si$
which are 
favorable experimentally, since they can be grown by sputtering
techniques~\cite{sprungmann}. 
The conducting ferromagnet $\rm CrO_2$~\cite{singh}
 is also a candidate for use in half-metallic spin valves, although 
it cannot be grown by sputtering methods, and is metastable.

%[Now discuss previous HM experiments].

%[fist, spin valve, tc]
Experimental signatures of triplet
 correlations 
in half-metallic 
$S F_1 F_2$  
spin valves
have been demonstrated  in 
 transition temperature  $T_c$ variations that occur when
rotating the magnetization  of the free ferromagnet layer. \cite{singh,robby}
Measuring the corresponding  maximal change
 in the critical temperature, $\Delta T_c$,
 can represent the emergence of spin-polarized triplet pairs 
 as the
 singlet superconducting state   weakens
and is subsequently converted into 
opposite-spin and equal-spin triplet pairs.
Most experiments for these types of 
spin valve structures involved weak ferromagnets for the outer $F_2$
layer and in-plane magnetic fields, yielding $\Delta T_c$ sensitivities 
from a few mK to around 100 mK. \cite{zhu,lek,wang,ilya,flok}
When the $F_2$ layer is replaced by a
the half-metallic ferromagnet such as $\rm CrO_2$, a  larger $\Delta T_c$
of $\Delta T_c \approx 800\,{\rm mK}$ was measured\cite{singh}
using  a large out-of-plane applied magnetic field. 
If  
$\rm La_{0.6} Ca_{0.4} MnO_3$ is used as the half metallic ferromagnet,  
a much weaker in-plane magnetic field suffices to rotate the magnetization in one of the $F$ layers~\cite{robby},
resulting  $\Delta T_c \approx  150\, \rm{mK}$, which again
is a stronger spin valve effect compared to 
experiments involving standard ferromagnets~\cite{lek,wang}. 
These types of improvements  were shown to be
consistent with theoretical work\cite{hmkh}
which   demonstrated  that when the exchange field in $F_2$ varies from
zero to half-metallic,  
the largest $\Delta T_c$ arises  when $F_2$ is 
a half-metallic. These experimental evidences further established %ct
the advantages of utilizing
half-metallic elements 
in superconducting spintronics devices.

%[Next,  tunnel junction and HM JJ: importance]

%JJ
Although critical temperature measurements give valuable  information regarding
half-metallic spin valves,
for spintronics devices it is important to also investigate the 
transport of 
charge and spin 
in these types of spin-valve structures.
By placing the spin valve between
two superconducting banks with a phase difference 
$\Delta\varphi$, a half-metallic based Josephson junction  with 
 spin-controlled supercurrent can be
generated.
Interest in Josephson 
junctions with ferromagnetic
 layers has grown due 
to their use in
cryogenic spintronic systems, 
including  
superconducting computers and nonvolatile memories,\cite{eshy,effy,golubov,buzdin1}
where their use in single flux quantum circuits 
can improve switching speeds.\cite{giaz,spath,ali}
 To determine whether Josephson structures can serve as 
 viable  spintronic devices, it is crucial to understand the 
 behavior of the spin currents that can flow in such systems. 
%The spin current that interacts with the magnetizations
The interaction between the spin currents and the magnetizations
in ferromagnetic Josephson junctions 
is important for memory applications since
the magnetization orientations in the $F$ layers dictates the
storage of information bits.
Controlling the magnetization 
rotation can be achieved by a  torque 
from
the  spin-polarized  
currents flowing perpendicular to the layers.
Some of the spin angular momentum of the polarized current 
will be transferred to the ferromagnets, giving rise to the STT %ct
 effect~\cite{slon, bergerl, linder, linder2, sac, sstech}.
 This effect can result in a
decrease of magnetization switching times in 
random access memories~\cite{fart, brataas_nat, Bauer_nat}.
The STT effect  is known to occur in a  broad variety
of ferromagnetic materials, including half-metals, making it 
widely accessible experimentally.

An essential mechanism responsible for supercurrent flow 
in a half-metallic Josephson junction
is Andreev reflection that occurs at
 the  ferromagnet and superconductor  interfaces.\cite{radovic2, radovic1, beenaker2, beenaker1}
 In addition to continuum  states, the superposition of 
localized quasiparticle wavefunctions in the ferromagnet regions results in 
subgap bound states that contribute to the total current flow.
For strong ferromagnets,
the corresponding spin-polarized Andreev
bound states can be strongly affected  by the supercurrent, directly
influencing the spin currents and STT when varying the relative in-plane magnetization
angle. 
Although the charge current is conserved, 
remaining uniform throughout the sample, the spin current
often varies spatially, making comparisons between  the two types of current difficult.
Moreover, since manipulating  
the angle
between the magnetization vectors can generate  long
ranged spin polarized triplet supercurrents~\cite{hvw15},
these triplet correlations
also correlate with spatial variations
in the spin currents responsible for the mutual torques
acting on the ferromagnets.

As demonstrated in Refs.~\onlinecite{robinson,khaire},
these equal-spin triplet pairs  result in 
a more robust Josephson
supercurrent that is relatively insensitive to $F$ layer thicknesses
due to their long-ranged nature.  %\cite{robinson,khaire}
If  one of the ferromagnets in the junction is  half-metallic, the equal-spin triplet correlations 
are expected to play an even greater role  
in the behavior of the charge and spin currents.
This was shown experimentally\cite{keizer} where a spin triplet supercurrent was measured through
the half-metal $\rm CrO_2$, and whose direction was switchable via magnetization variation. 
Even in the diffusive limit, it was shown that spin-flip scattering events at the interfaces of
a half-metallic Josephson junction also allow penetration of the equal-spin pairs into the half-metal~\cite{gobu}.
Considering the potentially greater control  of spin currents afforded by  Josephson 
junctions with strongly spin-polarized ferromagnets, it would be illuminating  to
systematically investigate the interplay of the triplet pair correlations with the charge and spin transport
throughout half-metallic Josephson structures.

%[Discuss tunneling method]
Another way to produce
 charge and spin currents in half-metallic spin valve 
 structures involves establishing 
 a voltage difference between the ends 
of a $F_1F_2S$ tunnel junction, resulting in an
injected current into the $F_1$ layer.
The charge and spin transport properties
for these types of nonequilibrium tunnel junctions with relatively weak ferromagnets
was previously studied~\cite{wvh14,mv17} as functions of bias voltage using 
a transfer matrix approach that combines the Blonder-Tinkham-Klapwijk (BTK) formalism and
self-consistent solutions to the Bogolibuov-de Gennes (BdG) equations. 
The use of this technique was also extended to accurately compute %kh
spin transport quantities, including  STT %kh the
and the spin currents,
%can be accurately determined
%, and 
while ensuring that the appropriate  conservation laws are satisfied. %kh
If the $F_1$ layer is half-metallic,
the current can become strongly polarized, leading to a relatively large
transfer of angular momentum to the $F_2$ layer for noncollinear 
magnetizations, via the %kh
STT effect.  
Also, the angularly averaged subgap conductance
 in this case  arises mainly from anomalous Andreev reflection~\cite{wvh14},
whereby 
a reflected hole with the same spin as the incident particle
is Andreev reflected,
generating a spin-polarized triplet pair.
The effects of applied bias on the spin transfer
torque and the spin-polarized tunneling conductance
has also been previously studied in superconducting   
tunnel junctions~\cite{linder0}.
By applying an 
external
magnetic field, 
or through switching via STT, it is again possible to %kh
control the relative orientation of the intrinsic magnetizations 
and investigate the dependence of the charge and spin currents
on the 
%orientation 
%misoriented  %kh
misorientation
angle $\theta$ between the
two ferromagnetic layers. 
Thus, when a half-metallic layer is present 
in a $F_1F_2S$ tunnel junction,
we can have
greater control
and isolation  of the
spin currents and spin-polarized 
triplet pairs 
that are critical for viable 
spintronics platforms.
The  systematic investigation into 
the transport  and corresponding  triplet correlations
of half-metallic spin valves
for both equilibrium Josephson junctions and
 nonequilibrium tunnel junctions 
is the 
%key point
main focus of this paper.

%[What we propose - how we do it]
When considering spin transport in superconducting junctions,
 it is beneficial for the structure  to contain both
weakly polarized and strongly polarized ferromagnets.
This is because 
the singlet and the opposite spin triplet correlations in 
weaker ferromagnets
extend over greater lengths, 
 dictated by the inverse of the exchange field,
 and
 they  are therefore  much  more
 effective at hosting
opposite-spin pairs.
The weak ferromagnet serves as an intermediate layer between the superconductor and
half-metal, facilitating  the generation of 
opposite-spin pairs that will eventually become  
converted  into  the
 longer ranged equal-spin
triplet pairs.
A hybrid ferromagnetic setup also creates an avenue for the   systematic investigation into
the interplay and ultimate control of
both triplet channels.
We therefore are interested in two types of tunnel junctions in this paper. 
The first consists of a single superconductor   in contact with
two ferromagnets (an $F_1 F_2 S$ structure), with the $F_2$ ferromagnet 
having a weak exchange field, and the other $F_1$,
half-metallic.
The current in this nonequilibrium case is injected by means of a voltage difference between two electrodes.
As alluded to earlier,
the other scenario involves 
 a Josephson junction containing 
a half-metal flanked by two weaker conventional ferromagnets.
The current  is established in the usual way by a macroscopic phase difference
$\Delta\varphi$ between  the  two outer superconducting  banks. 
For both junction arrangements, we investigate the charge
and spin transport within the ballistic 
regime using a microscopic self-consistent 
BdG formalism
 that is capable of accommodating the broad range of energy scales set by
the  
exchange field $h$ of the
conventional  ferromagnets ($h/E_F\ll 1$)
and the half-metal ($h=E_F$).
Of crucial importance towards the theoretical description of these 
type of transport structures is to
accurately be able to  account for the mutual interactions between 
the ferromagnetic and superconducting 
elements, i.e., 
proximity effects.
This requires a self-consistent treatment, which  
ensures that 
the final solutions minimize the free energy of the 
system and satisfies the proper conservation laws.
This numerical approach  is a time-consuming but necessary 
step to
reveal  the self-content 
 proximity
effects that govern the nontrivial
charge and spin currents that flow within these structures.
Indeed,
the tunneling conductance  in $F_1F_2S$ junctions  was shown to 
differ substantially from that obtained
via a non-self-consistent approach~\cite{wvh14}.

%Details of equilibrium approach. Details of non-equilibrium approach. [copy from earlier two papers]

This paper is organized as follows: In Sec.~\ref{description},
we present the general  Hamiltonian and self-consistent
BdG methodology that is applicable for both junction configurations.
In Sec.~\ref{BTK}, the
transfer matrix approach
for tunnel junctions that combines 
the Blonder-Tinkham-Klapwijk (BTK) formalism and
self-consistent solutions to the BdG equations is established. %ct
The charge continuity equation and current density are also 
derived.
% both zero 
%and
% finite bias.
In Sec.~\ref{jojo}, the relevant details for the characterization of equilibrium
half-metallic Josephson junctions and
the expression for the associated current density are given. 
%Also the charge and current density
%are presented for equilibrium situations.
In Sec.~\ref{tripcorr}, we outline how to calculate the induced 
triplet correlations for equilibrium Josephson junctions %kh
and non-equilibrium tunnel junctions.
In Sec.~\ref{spintrans}
%, as well as 
the techniques used to
compute the spin transport quantities including
%charge current, 
magnetization, spin-transfer torque,  and the spin current
are derived for both types of junctions.
Throughout Sec.~\ref{methods}, we discuss
how to properly satisfy the conservation laws for
charge and spin densities in our formalism.
%Here we also outline how the
%triplet correlations are calculated using our method. 
In Sec.~\ref{tj} we present the results for half-metallic tunnel junctions. 
Results for the spatial dependence to the bias-induced magnetizations, 
the spin-transfer torque,  the spin currents, and triplet correlations 
are presented as functions of the magnetization misalignment angle 
as well as the applied bias.
We also report how to take advantage of the induced triplet correlations
by choosing the optimal exchange interactions in $F$ layers.
In Sec.~\ref{jojo_results}, we present the results for the half-metallic Josephson junctions,
including the current phase relations for a variety of half-metal thicknesses.
The spatial dependencies to the spin currents and triplet correlations are given,
and a broad range of misalignment angles are considered to demonstrate the
propagation of spin-polarized triplet pairs through the half-metal.
The positive correlations between the equal-spin triplet correlations
and the spin-polarized supercurrents are also discussed. 
We conclude with a summary in Sec.~\ref{conclusion}.

\section{Methods}
\label{methods}
\subsection{Description of the systems}
\label{description}
Two types of half-metallic junctions are considered in this paper:
%Both the 
tunneling junctions and Josephson junctions. The effective Hamiltonian that is applicable to both types of junctions is 
\begin{eqnarray}
\label{ham}
{\cal H}_{\rm eff}&=&\int d^3r \left\{ \sum_s
\psi_s^{\dagger}\left(\bm{r}\right)H_0
\psi_s\left(\bm{r}\right)\right.\nonumber \\
&+&\left.\frac{1}{2}\left[\sum_{s\:s'}\left(i\sigma_y\right)_{ss'}
\Delta\left(\bm{r}\right)\psi_s^{\dagger}
\left(\bm{r}\right)\psi_{s'}^{\dagger}
\left(\bm{r}\right)+H.c.\right]\right.\nonumber \\
&-&\left.\sum_{s\:s'}\psi_s^{\dagger}
\left(\bm{r}\right)\left(\bm{h}\cdot\bm{\sigma}
\right)_{ss'}\psi_{s'}\left(\bm{r}\right)\right\},
\end{eqnarray}
where $H_0$ is the single-particle part of ${\cal H}_{\rm eff}$, $\bm{h}$
describes exchange interaction of the magnetism, $s$ and $s'$ are spin indices
and $\bm{\sigma}$ are Pauli matrices.
$\Delta(\bm{r})\equiv 
g\left(\bm{r}\right)\left\langle\psi_{\uparrow}
\left(\bm{r}\right)\psi_{\downarrow}\left(\bm{r}\right)  
\right\rangle$ is 
the superconducting pair potential and $g(\bm r)$ is the coupling constant.
In ferromagnets where there is no
intrinsic superconducting pairing, $g(\bm r)$ is taken to be zero.
Similarly, $\bm{h}$ vanishes in intrinsically superconducting regions.
Following Ref.~\onlinecite{wvh14}, we utilize
the generalized Bogoliubov transformation~\cite{bdg}, 
$\psi_s=\sum_n\left(u_{ns}\gamma_n
+\eta_sv_{ns}^{\ast}\gamma_n^{\dagger}\right)$,
where  $\eta_s\equiv1(-1)$ for spin-down (up),
to write down the BdG Hamiltonian equivalent to Eq.~(\ref{ham}):
\begin{align}
&\begin{pmatrix}
H_0 -h_z&-h_x+ih_y&0&\Delta \\
-h_x+ih_y&H_0 +h_z&\Delta&0 \\
0&\Delta&-(H_0 -h_z)&-h_x+ih_y \\
\Delta&0&-h_x+ih_y&-(H_0+h_z) \\
\end{pmatrix}
\begin{pmatrix}
u_{n\uparrow}\\u_{n\downarrow}\\v_{n\uparrow}\\v_{n\downarrow}
\end{pmatrix} \nonumber \\
&=\epsilon_n
\begin{pmatrix}
u_{n\uparrow}\\u_{n\downarrow}\\v_{n\uparrow}\\v_{n\downarrow}
\end{pmatrix}\label{bogo},
\end{align}
where  $u_{ns}$ and $v_{ns}$ in the
generalized Bogoliubov transformations 
can be identified as  the quasiparticle 
and quasihole amplitudes, respectively. 

For layered tunnel junctions 
and Josephson junctions 
considered in this work,
we assume each $F$ and $S$ layer 
is infinite in the $yz$ plane %kh
and the layer thicknesses extend 
along the $x$ axis (See Figs.~\ref{ffs_struct} and~\ref{diagram}). 
As a result, the BdG Hamiltonian [Eq.~(\ref{bogo})] is translationally invariant
in the $yz$ plane, and it becomes quasi-one-dimensional in $x$. %kh
The single-particle Hamiltonian is
$H_0=-({1}/{2m})({d^2}/{dx^2})+{\epsilon_{\perp}}-E_F$, where
we have defined the transverse kinetic energy as 
$\epsilon_{\perp}\equiv (k_y^2+k_z^2)/2m$,
and  $E_F$ denotes the Fermi energy. Although in this work, we do not consider  %kh
Fermi energy mismatch between distinct layers, it is straightforward
to include such an effect.
Throughout this paper, we take $\hbar=k_B=1$, and all energies
are measured in units of $E_F$. We numerically determine the pair potential 
by using fully self-consistent solutions
to Eq.~(\ref{bogo}). The iterative self-consistent procedure 
has been extensively discussed in previous work~\cite{wvh12,wvh14}.
Since our BdG Hamiltonian is quasi-one-dimensional,
the pair potential is only a function of $x$. 
By minimizing the free energy of the system, and %kh
making use of the generalized
Bogoliubov transformation, the pair potential  can be written as, %kh
\begin{equation}
\label{del}
\Delta(x) = \frac{g(x)}{2}{\sum_n}^\prime
\bigl[u_{n\uparrow}(x)v_{n\downarrow}^{\ast}(x)+
u_{n\downarrow}(x)v_{n\uparrow}^{\ast}(x)\bigr]\tanh\left(\frac{\epsilon_n}{2T}\right), \,
\end{equation}
where $T$ is temperature and the prime symbol means that a Debye cutoff energy, $\omega_D$, 
is introduced in the energy sum.
Additional details of our formalism used in this work 
can also be found in Refs.~\onlinecite{wvh14,hvw15}. 
%The superconducting coupling
%strength $g(x)$ is a nonzero constant on the S side and vanishes
%in the F regions. 

\subsection{Tunnel junctions}
\label{BTK}
\begin{figure}
\includegraphics
[width=3.2in]
{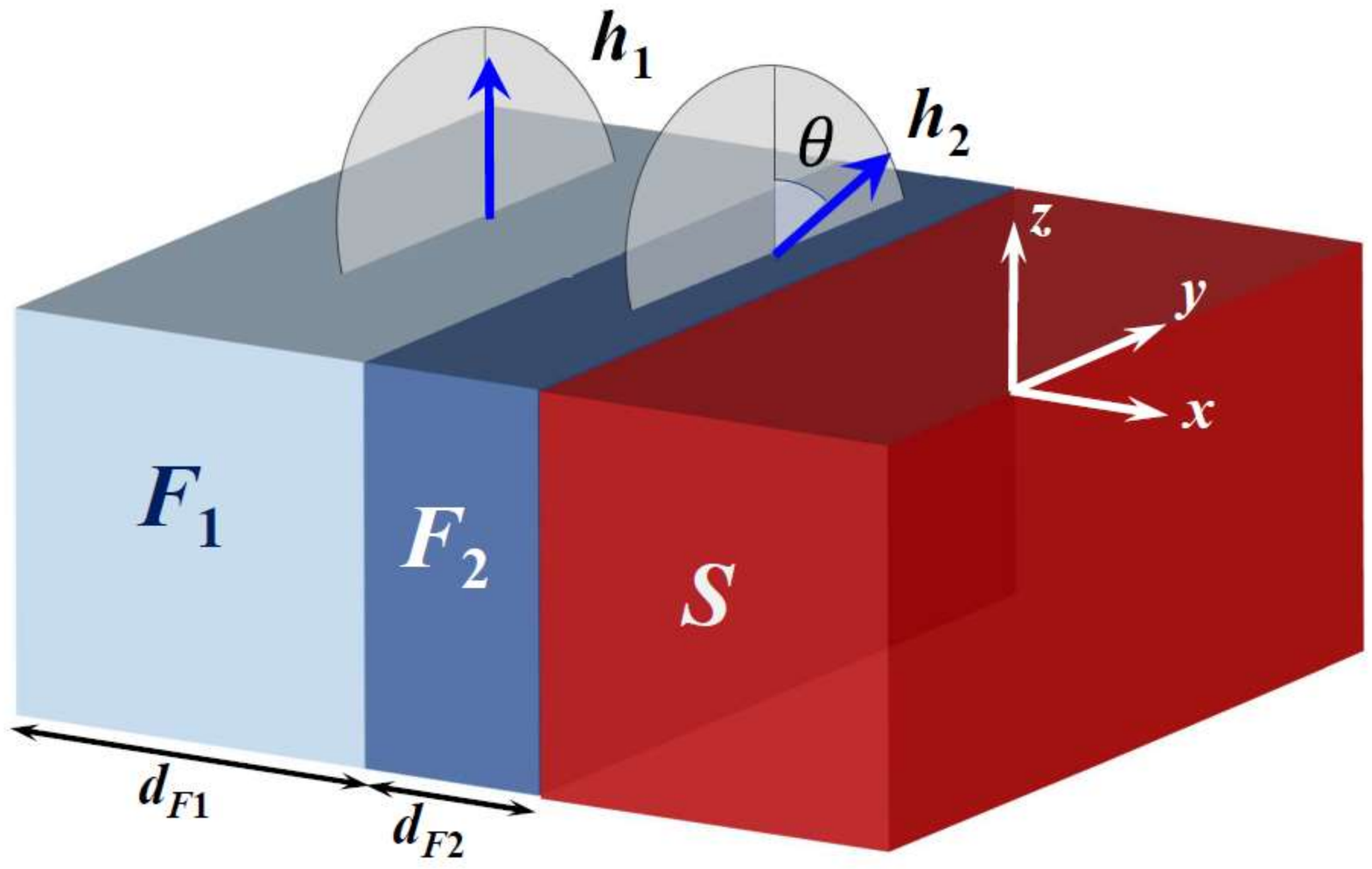}
\caption{
(Color online) Illustration of the ${F_1F_2S}$ %kh
tunnel junction that  is infinite and translationally %kh
invariant in the $yz$ plane. It has finite 
size along the $x$ axis. 
$F_1$ is a half-metal and the %kh
associated exchange field is fixed along
the $z$ direction. %ct
The direction of the exchange field in
$F_2$ is in the $yz$ plane, and %kh
makes an angle $\theta$ with the $z$
axis. Such a misorientation can
be achieved experimentally
via an external magnetic field.
}
\label{ffs_struct}
\end{figure}

We begin first with tunnel junctions depicted in Fig.~\ref{ffs_struct}
where a ferromagnet and half metal are in contact with a superconductor.
The ferromagnet that
is not adjacent to $S$ is labeled  $F_1$, and the one next to $S$
is $F_2$. As shown in Fig.~\ref{ffs_struct},
the exchange field in $F_1$ is
$h_1\hat{\bm z}$, and in $F_2$
it is $h_2\left(\sin\theta\hat{\bm y}+\cos\theta\hat{\bm z}\right)$.
Here $h_1$ and $h_2$ are  the magnitudes of the exchange fields in $F_1$
and $F_2$, respectively. 
In general, we consider $F_1$
as a fixed layer where the exchange field is pinned
 and $F_2$ as a free layer where the relative angle
$\theta$ can be controlled by an applied magnetic field experimentally~\cite{ilya}. 
In this work, we take the fixed layer $F_1$ to be a half metal and $h_1=E_F.$

In  previous work~\cite{wvh14}, a formalism based on the BTK approach~\cite{btk}
was  generalized to study  spin-transport quantities.
In Ref.~\onlinecite{btk}, it was shown, starting from the Boltzmann equation, %kh
that the conductance associated with the tunnel junction 
is a function of the transmission and reflection amplitudes %kh
in the linear response regime. Therefore, to compute the  tunneling conductance,
one should start by writing down the appropriate wavefunctions in each distinct %kh
region. If one considers a bilayer tunnel junction that is %kh
made up of a non-magnetic metal and a superconductor, then the eigenfunctions
in the non-magnetic metallic region are only  linear combinations of
particle and hole wavefunctions.\cite{btk}  %kh
%as in Ref.~\onlinecite{btk}. 
However,
in our work, where the non-magnetic metallic region
is replaced by two ferromagnetic layers, one should consider
the spin degree of freedom 
%on top of 
in addition to the particle-hole nature. %kh
Because the exchange field is along $z$ in $F_1$,
the appropriate eigenfunctions are
\begin{equation}
\label{f1wave}
\begin{pmatrix}
e^{\pm ik^+_{\uparrow1}x}
\\0
\\0
\\0
\end{pmatrix},\enspace
\begin{pmatrix}
0
\\e^{\pm ik^+_{\downarrow1}x}
\\0
\\0
\end{pmatrix},\enspace
\begin{pmatrix}
0
\\0
\\e^{\pm ik^-_{\uparrow1}x}
\\0
\end{pmatrix},\enspace
\begin{pmatrix}
0
\\0
\\0
\\e^{\pm ik^-_{\downarrow1}x}
\end{pmatrix},
\end{equation}
where the subscript 1 denotes the $F_1$ regions
and the superscript $+$ is for particle-like and $-$ is for hole-like wavefunctions.
When the eigenenergy $\epsilon$ is specified, the corresponding wavevectors
are given by the following relation
\begin{equation}
\label{momentum}
k_{s1}^\pm=\left[1-\eta_s h_1\pm \epsilon -k_\perp^2\right]^{1/2},
\end{equation}
where $k_\perp^2=k_y^2+k_z^2$.
The incident angle, $\theta_I$, relative to the normal of the interface 
with spin $s$ 
is related to $k_\perp$ and given by the relation,
$\tan\theta_I=k_\perp/k_{s1}^\pm$.
The reflected angles, $\theta_R$, similarly obey $\tan\theta_R=k_\perp/k_{s1}^\pm$.
From Eq.~(\ref{momentum}), it is easy to see that the  reflected angles
depend on both the spin as well as 
whether the quasiparticle  is  particle-like or hole-like.
The exchange field in ${F_2}$ lies on the $yz$  plane,   
and it is tilted relative to the $z$-axis by the angle $\theta$.
One needs again to use suitable eigenfunctions for both 
particle and hole branches in ${F_2}$. 
The particle-like wavefunction with spin
parallel to the exchange field in ${F_2}$
and antiparallel to the exchange field in ${F_2}$ are given as
\begin{equation}
\label{f2spinor1}
\begin{pmatrix}\cos\left(\theta/2\right)\\\sin\left(\theta/2\right)\\0\\0\end{pmatrix}
e^{\pm ik^+_{\uparrow2}x},\enspace\enspace
\begin{pmatrix}-\sin\left(\theta/2\right)\\\cos\left(\theta/2\right)\\0\\0\end{pmatrix}
e^{\pm ik^+_{\downarrow2}x},\enspace\enspace
\end{equation}
respectively.
Similarly, the hole-like wavefunction with spin
parallel and antiparallel to the exchange field in $F_2$ are 
given by
\begin{equation}
\label{f2spinor2}
\begin{pmatrix}0\\0\\\cos\left(\theta/2\right)\\-\sin\left(\theta/2\right)\end{pmatrix}
e^{\pm ik^-_{\uparrow2}x},\enspace\enspace
\begin{pmatrix}0\\0\\\sin\left(\theta/2\right)\\\cos\left(\theta/2\right)\end{pmatrix}
e^{\pm ik^-_{\downarrow2}x},\enspace\enspace
\end{equation}
respectively.
Here the momenta are defined through the relation
\begin{equation}
k_{s2}^\pm=\left[1-\eta_s h_2\pm \epsilon -k_\perp^2\right]^{1/2}.
\end{equation}
Note here that following previous conventions, we denote $``+"$ 
for particles, and $``-"$ for holes.
Because the Hamiltonian is translationally invariant in the $yz$ %kh
plane, the perpendicular momentum $k_\perp$ is a constant 
throughout the ``entire" junction for a given eigenstate appropriate to
the entire junction. 
Once the energy of the eigenstate, $\epsilon$, is prescribed, 	
the eigenfunctions in the $F_2$ region are given as a %kh
linear combination of these wavefunctions. 
Accordingly,
there are eight unknowns associated with this linear combination.
On the superconducting side, one can easily show that in $4\times4$
Nambu space, the appropriate wavefunctions are
\begin{equation}
\label{Swave}
\begin{pmatrix}
u_0
\\0
\\0
\\v_0
\end{pmatrix}e^{\pm ik^+x},\enspace
\begin{pmatrix}
0
\\u_0
\\v_0
\\0
\end{pmatrix}e^{\pm ik^+x},\enspace
\begin{pmatrix}
v_0
\\0
\\0
\\u_0
\end{pmatrix}e^{\pm ik^-x},\enspace
\begin{pmatrix}
0
\\v_0
\\u_0
\\0
\end{pmatrix}e^{\pm ik^-x},
\end{equation}
where $k^{\pm}=1\pm\sqrt{\epsilon^2-\Delta_0^2-k_\perp^2}$.
If a non-self-consistent pair potential is adopted for which %kh
the pair potential in the $S$ region is a constant, the entire
$S$ region is just a linear combination of the above wavefunctions
with suitable constants $u_0$ and $v_0$ given by,
\begin{subequations}
\begin{align}
u_0^2=&\frac{1}{2}\left(1+\frac{\sqrt{\epsilon^2-\Delta_0^2}}{\epsilon}\right),\\
v_0^2=&\frac{1}{2}\left(1-\frac{\sqrt{\epsilon^2-\Delta_0^2}}{\epsilon}\right),
\end{align}
\end{subequations}
where $\Delta_0$ is the constant pair amplitude.
Let us first discuss the non-self-consistent case and
suppose a spin-up particle is sent from an electrode into the $F_1$ region.
In the $F_1$ region, one needs to include the incident spin-up 
particle wavefunction as well as four different types of reflection:
(1) a reflected particle wavefunction with spin-up, %kh added 'a'
(2) a reflected particle wavefunction with spin-down,
(3) an Andreev reflected hole wavefunction with spin-up,  %ct
and (4) an Andreev reflected hole wavefunction with spin-down. %ct
As a result, we have four unknowns associated with these 
four reflected wavefunctions.
In the $F_2$ region, all eight possibilities, Eqs.~(\ref{f2spinor1})
and (\ref{f2spinor2}), must be considered, since in general 
the waves can travel in either the $+x$ or $-x$ directions.
In the $S$ region, there are four different types of transmitted wavefunctions:
two transmitted particle-like wavefunctions,
\begin{equation}
\begin{pmatrix}
u_0
\\0
\\0
\\v_0
\end{pmatrix}e^{ik^+x},\enspace
\begin{pmatrix}
0
\\u_0
\\v_0
\\0
\end{pmatrix}e^{ik^+x},\enspace
\end{equation}
and two transmitted hole-like wavefunctions,
\begin{equation}
\label{Swave}
\begin{pmatrix}
v_0
\\0
\\0
\\u_0
\end{pmatrix}e^{-ik^-x},\enspace
\begin{pmatrix}
0
\\v_0
\\u_0
\\0
\end{pmatrix}e^{-ik^-x}.
\end{equation}
Thus, the total number of unknowns in this process is sixteen
(four from  the reflections, eight associated with the $F_2$ region,
and four from the transmissions). 
We have exactly the same number of constraints to solve for these 
unknowns because there are two interfaces ($F_1/F_2$ and 
$F_2/S$) at which the continuous conditions of the wavefunction and
its derivative must hold when the interfacial
barrier is absent. 

If one uses a self-consistent profile for the pair amplitude,
$\Delta$ is not a constant and it varies with $x$. %ct
It is convenient to consider a transfer-matrix approach 
to take into account the variation of $\Delta$.
The details of this approach are presented in Ref.~\onlinecite{wvh14}
and will not be repeated in this paper. Here, we only summarize
the outline of this approach.
One first divides the $S$ region into a number of small subregions %kh S
and approximates each subregion by a constant potential.
One can then write down suitable wavefunctions in each subregion. 
Except for the last subregion where there are only four unknowns linked to
four types of transmission, there are eight unknowns 
associated with each subregion, resulting  now in an overall
greater number of unknowns. 
%To solve this 
%effectively, we employ a useful transfer matrix method~\cite{wvh14}.
By recognizing the fact that unknowns on one side of an interface
are related to those on the other side, we can write,
\begin{equation}
\tilde{\mathcal{M}}_ix_i=\mathcal{M}_{i+1}x_{i+1},
\end{equation}
where $i$ is the index of each subregion,  
$\tilde{\mathcal{M}}_i$ and $\mathcal{M}_{i+1}$ are the
corresponding matrices determined by matching the boundary conditions,
and $x_i$ and $x_{i+1}$ are the column vectors composed of the unknowns
in the $i$-th and $i+1$-th subregions.
By using this recurrence relation, one naturally relates
the reflection coefficients in the $F_1$ region with 
the transmission coefficients in the outermost $S$ layer.
Once these transmission and reflection coefficients
are found, they can be fed back into the recurrence relation
to  generate solutions in each subregion. 
The transfer matrix method is advantageous because the size of the 
matrix equation needed to be solved is much smaller than 
the number of unknowns, albeit at the cost of
multiplying matrices.

The BTK formalism was originally developed to 
extract the tunneling conductance from transmitted
and reflected amplitudes. The formula for 
spin-dependent conductance,
normalized to that of the normal state, in the low
temperature regime is given by
\begin{equation}
\label{conductance}
G_s=1+\frac{k^-_{\uparrow 1}}{k^+_{s1}}|a_{s\uparrow }|^2
+\frac{k^-_{\downarrow 1}}{k^+_{s1}}|a_{s\downarrow }|^2
-\frac{k^+_{\uparrow 1}}{k^+_{s1}}|b_{s\uparrow }|^2
-\frac{k^+_{\downarrow 1}}{k^+_{s1}}|b_{s\downarrow }|^2,
\end{equation} 
where $a_{s\uparrow}$ and $a_{s\downarrow}$ are Andreev
reflected waves and $b_{s\uparrow}$ and $b_{s\downarrow}$
are normal reflected waves.
In the above expression, the subscript $s$ denotes
the spin type of the incident wave into the $F_1$ region.

In Ref.~\onlinecite{wvh14}, the BTK formalism has been generalized
to study transport quantities such as spin currents
and spin transfer torques. 
By applying the transfer matrix method outlined above, 
these position dependent quantities can be properly computed.
Below, we shall describe the basic ideas behind our approach.
From the Heisenberg equation for the charge density $\rho({\bm r})$,
\begin{equation}
\label{heisenberg}
\frac{\partial}{\partial t}\left\langle\rho({\bm r})\right\rangle
=i\left\langle\left[{\cal H}_{eff},\rho({\bm r})\right]\right\rangle,
\end{equation}
it is not difficult to obtain the following continuity condition
for the current density ${\bm J}$:
\begin{equation}
\frac{\partial}{\partial t}\left\langle\rho(\bm r)\right\rangle
+\nabla\cdot{\bm J}=
-4e{\rm Im}\left[\Delta({\bm r})\left\langle 
\psi_{\uparrow}^{\dagger}({\bm r})
\psi_{\downarrow}^{\dagger}({\bm r})\right\rangle\right].
\label{current}
\end{equation}
When in the steady state, the first term on the left is dropped.
Moreover, when the system is in equilibrium without an external bias,
one can use the Bogoliubov transformation together with
the conservation law for our quasi-one-dimensional system
to conveniently write the continuity equation as:
\begin{equation}
\frac{\partial J_x(x)}{\partial x}= 2e {\rm Im}\left\{\Delta(x)\sum_n\left[u_{n \uparrow}^* 
v_{n \downarrow}+u_{n\downarrow}^*v_{n\uparrow}\right]\tanh\left(\frac{\epsilon_n}{2T}\right)\right\}.
\label{currentuv}
\end{equation}
The self-consistency condition, Eq.~(\ref{del}),
demands that the right hand side of Eq.~(\ref{currentuv}) %kh
vanishes and that the current is a constant throughout the junction, as expected. %kh
%ct Note here that the above discussion is appropriate for both Josephson junctions
%and tunnel junctions.

Now consider a non-zero bias, $V$, across the electrodes
of a $F_1F_2S$ junction. 
The bias generates  a non-equilibrium quasi-particle distribution. %kh
In the excitation picture, 
it is clear that all states with energies $\epsilon<eV$ incident from 
the electrode in $F_1$ to the electrode in $S$ should be taken into
account in the low $T$ limit~\cite{wvh14}. Hence,
the charge density and the current density can be
derived, and are given by:
\begin{align}
\rho&=-e\sum_{ns}|v_{ns}|^2
-e\sum_{\epsilon_{\bm k}<eV}\sum_s\left(|u_{{\bm k}s}|^2-|v_{{\bm k}s}|^2\right),\\
J_x&=-\frac{e}{m}{\rm Im}\left[
\sum_{\epsilon_{\bm k}<eV}\sum_s\left(u_{{\bm k}s}^{\ast}\frac{\partial u_{{\bm k}s}}{\partial x}
+v_{{\bm k}s}^{\ast}\frac{\partial v_{{\bm k}s}}{\partial x}\right)\right],
\label{current2}
\end{align}
where we sum over states labeled by their momenta $\bm{k}$ with energies less than the bias.
It is easy to see from the above equations that when $V=0$, $J_x=0$, and 
$\rho$ is just the ground-state charge density, as one would expect.
The right hand side of the continuity equation, Eq.~(\ref{current}),
with the presence of the bias, becomes
$-4e{\rm Im}\left[\Delta\sum_{\epsilon_{\bm k}<eV}\left(u_{{\bm k}\uparrow}^{\ast}
v_{{\bm k}\downarrow}+v_{{\bm k}\uparrow}u_{{\bm k}\downarrow}^{\ast}\right)\right]$.
We emphasize here that $\Delta$ vanishes in the intrinsically non-superconducting region since
the coupling constant is taken to be zero there. Hence, on the $F$ side %ct
the spatial derivative of the current vanishes and the current is a constant.
On the $S$ side, where $\Delta$ exists, the derivative of the current does not vanish.
This does not mean that the conservation law is violated. 
The right-hand-side actually describes the process of interchange
between the quasi-particle current density and
the supercurrent density, as clearly discussed in Ref.~\onlinecite{btk,wvh14}.

\subsection{Josephson junctions}
\label{jojo}
We next discuss the pertinent aspects of the half-metallic 
Josephson junctions that we shall investigate.
As shown in Fig.~\ref{diagram},
we consider $S_1F_1F_2 F_3 S_2$ type  junctions,
where the central half-metallic layer $F_2$ is surrounded by two 
ferromagnets $F_1$ and $F_3$.  We will show below in Sec.~\ref{jojo_results} that it is important %ct
for the ferromagnets to be thin (relative to $\xi_F$, the superconducting proximity length)
and for them  to have relatively weak exchange fields
so that their placement near the superconducting banks allows for the generation of triplet
correlations and the associated phase coherent transport.
The exchange fields in each of the junction layers reside
in-plane and
are written
\begin{align} \label{have}
\bm h_i=h_i(\sin\theta_i \hat{\bm y} + \cos\theta_i \hat{\bm z}),\quad \text{for $i=1,2,3$}.
\end{align}
To compute the dc Josephson current where the bias 
across the junction is absent, we again 
numerically look for solutions by iteratively solving Eq.~(\ref{bogo}),
which is very general and can be applied to both the $F_1F_2S$ tunneling %kh
and $S_1F_1F_2S_2$ Josephson junctions.
Since we wish to determine the current-phase relation for 
the Josephson junctions, the initial input for the pairing potential 
is taken to be the bulk gap, $\Delta_0$, in ${S_1}$
and $\Delta_0 \exp(i\Delta\varphi)$ in $S_2$.
With this input, Eq.~(\ref{bogo}) is then numerically
diagonalized and the new pair potential, $\Delta(x)$ 
is computed from Eq.~(\ref{del}) throughout the entire junction except
for small regions (around one coherence length, $\xi_0$, from  the  sample %is it still the same? 
edges) considered as boundaries of the  junctions.
In these regions, the pair potential is fixed to its bulk
absolute value, with phases $0$ and $\Delta\varphi$, respectively.
The newly yielded $\Delta(x)$ is then used  in the BdG equations 
and the above  
process is repeated iteratively until convergence is achieved.
From Eq.~(\ref{current}),
when current is flowing through the junction, the self-consistently 
calculated regions
are always found to possess the necessary spatially constant current.
The important distinction between tunneling and Josephson junctions is 
the presence of the external bias. For dc Josephson junctions, the bias
is absent and the right-hand side of Eq.~(\ref{current}) should always
vanish in order to not  violate the conservation law. 
One can also write down the charge supercurrent associated with
a fixed nonzero phase difference between $S_1$ and $S_2$.
The expression for the current density in a Josephson junction is given by
\begin{align}
J_x&=-\frac{e}{m}\sum_{ns} {\rm Im }\left[u_{ns} \frac{\partial u^{*}_{ns}}{\partial x} f_n+
v_{ns}\frac{\partial v^{*}_{ns}}{\partial x} \left(1-f_n\right) \right],
\label{jj}
\end{align}
where $f_n$ is the Fermi function.
If the phase of the order parameter is a constant throughout the junction, 
the current density vanishes as can be seen from Eq.~(\ref{jj}).
We emphasize here that Eq.~(\ref{jj}) is applicable only when the external
bias is absent. Nevertheless, both Eqs.~(\ref{current2}) and (\ref{jj})
are derived using  the Heisenberg approach. 

\subsection{Triplet correlations}
\label{tripcorr}
As discussed in the introduction, for half-metallic superconducting
junctions, the induced spin-triplet Cooper pairs play an important role
in both equilibrium and transport properties.
These triplet pairing correlations are defined as
%\begin{subequations}
%\label{pa}
%\begin{align}
%{f_0}({\bm r},t) =& \frac{1}{2}[\langle \psi_{\uparrow}({\bm r},t) \psi_{\downarrow} 
%({\bm r},0)\rangle+
%\langle \psi_{\downarrow}({\bm r},t) \psi_{\uparrow} ({\bm r},0)\rangle],\\
%{f_1}({\bm r},t) =& \frac{1}{2}[\langle \psi_{\uparrow}({\bm r},t) \psi_{\uparrow} 
%({\bm r},0)\rangle -\langle \psi_{\downarrow}({\bm r},t) \psi_{\downarrow} 
%({\bm r},0)\rangle],
%\end{align}
%\end{subequations}
\begin{subequations}
\label{pa}
\begin{align}
{f_0}({\bm r},t) = &\frac{1}{2}\left[\left\langle
\psi_{\uparrow}({\bm r},t) \psi_{\downarrow} ({\bm r},0)\right\rangle+
\left\langle \psi_{\downarrow}({\bm r},t) \psi_{\uparrow} ({\bm r},0)\right\rangle\right],\\
\label{f1}
{f_1}({\bm r},t) = &\frac{1}{2}\left[\left\langle
\psi_{\uparrow}({\bm r},t) \psi_{\uparrow} ({\bm r},0)\right\rangle
-\left\langle \psi_{\downarrow}({\bm r},t) \psi_{\downarrow} ({\bm r},0)\right\rangle\right], \\
\label{f2}
{f_2}({\bm r},t) = &\frac{1}{2}\left[\left\langle
\psi_{\uparrow}({\bm r},t) \psi_{\uparrow} ({\bm r},0)\right\rangle
+\left\langle \psi_{\downarrow}({\bm r},t) \psi_{\downarrow} ({\bm r},0)\right\rangle\right],
\end{align}
\end{subequations}
where the subscript $0$  corresponds to $m_s=0$, and 
the subscripts $1$ and $2$ refer to the
$m_s=\pm1$ projections on the spin quantization axis. 
It was shown
in previous work that using this approach to find both the opposite-spin and equal-spin triplet pairs,
satisfies  the Pauli exclusion principle, and that the triplet pairs
vanish at $t=0$~\cite{hbvprl,hvb08}. If the exchange fields between in $F$ layers 
are not collinear, or equivalently, $\theta_i\neq0$, 
the total spin operator of the pairs does not commute with the effective Hamiltonian [Eq.~(\ref{ham})], and
the long-ranged, spin-polarized components $f_1$ and $f_2$ 
can be induced~\cite{hbvprl,hvb08}.
By using the generalized Bogoliubov transformation
and the Heisenberg equations
of motion,
it is possible to 
write the field operators in Eqs.~(\ref{pa}) as,
%in terms of quasi-particle and quasi-hole functions, we arrive at 
%If we consider the quantization axis fixed along the $z$ axis, the triplet amplitudes, $f_{0}(x,t)$ and $f_{1}(x,t)$, can be 
%written in terms of the quasiparticle amplitudes:~\cite{halter_trip,halter_trip2}
\begin{subequations}
\label{fall}
\begin{align}
f_{0}(x,t) &=  \frac{1}{2}\sum_{n}
\left[u_{n \uparrow}(x) v^{\ast}_{n\downarrow}(x)
-u_{n \downarrow}(x) v^{\ast}_{n\uparrow}(x)
%f_n^{\uparrow\downarrow}(x)-f_n^{\downarrow\uparrow}(x) %kh no need for abbreviation
\right] \zeta_n(t), \label{f0} \\
f_{1}(x,t) & =-\frac{1}{2} \sum_{n}
\left[
u_{n \uparrow}(x) v^{\ast}_{n\uparrow}(x)
+u_{n \downarrow}(x) v^{\ast}_{n\downarrow}(x)
%f_n^{\uparrow\uparrow}(x)+f_n^{\downarrow\downarrow}(x)
\right]\zeta_n(t), \label{f1} \\
f_{2}(x,t) & =-\frac{1}{2} \sum_{n}
\left[
u_{n \uparrow}(x) v^{\ast}_{n\uparrow}(x)
-u_{n \downarrow}(x) v^{\ast}_{n\downarrow}(x)
%f_n^{\uparrow\uparrow}(x)+f_n^{\downarrow\downarrow}(x)
\right]\zeta_n(t), \label{f2}
\end{align}
\end{subequations}
%\begin{subequations}
%\label{f0}
%\begin{align}
%f_{0} &=  \frac{1}{2}\sum_{n}(g_n^{\uparrow\downarrow}-g_n^{\downarrow\uparrow}) \zeta_n(t), \\
%\label{f1}
%f_{1} &=  \frac{1}{2}\sum_{n} (g_n^{\uparrow\uparrow}+g_n^{\downarrow\downarrow})\zeta_n(t),
%\end{align}
%\end{subequations}
where $\zeta_n(t) \equiv \cos(\epsilon_n t)-i\sin(\epsilon_nt)\tanh(\epsilon_n/2 T)$ 
and we have assumed zero bias for the junctions.
The triplet amplitudes in Eqs.~(\ref{f0})-(\ref{f2})
pertain to a fixed quantization axis along the $z$-direction.
In situations where it is more convenient to align the spin quantization axis with
the local magnetization direction, we rotate it using the transformations in 
the Appendix.
The exchange field orientations in each layer are described by the angle $\theta_i$,
and thus
we  write,
\begin{subequations}
\label{frot}
\begin{align}
f_0'&= \cos\theta_i f_0
+i\sin\theta_i f_2,\\
f_1'&=f_1,\\
f_2'&=\cos\theta_i f_2
+i\sin\theta_i f_0,
\end{align}
\end{subequations}
where the prime denotes the rotated system.

The triplet correlations given in Eqs.~(\ref{fall}) are only
applicable to both static and dynamic equilibrium situations
when the external bias is absent.
When $V\neq0$ and in the limit $T\rightarrow0$, 
Eqs.~(\ref{pa}) are bias dependent and we have
the following contributions in addition to Eqs.~(\ref{fall}),
\begin{subequations}
\label{findV}
\begin{align}
\delta f_{0}(x,t) &=  2i\sum_{\epsilon_{\bm k}<eV}
\left(u_{{\bm k}\uparrow}(x)v^{\ast}_{\bm k\downarrow}(x)-u_{{\bm k}\downarrow}(x)v_{{\bm k}\uparrow}^{\ast}(x)\right)\sin\left(\epsilon_{\bm k}t\right), \label{indf0} \\
\delta f_{1}(x,t) &=  2i\sum_{\epsilon_{\bm k}<eV}
\left(u_{{\bm k}\uparrow}(x)v^{\ast}_{\bm k\uparrow}(x)+u_{{\bm k}\downarrow}(x)v_{{\bm k}\downarrow}^{\ast}(x)\right)\sin\left(\epsilon_{\bm k}t\right), \label{indf1} \\ 
\delta f_{2}(x,t) &=  2i\sum_{\epsilon_{\bm k}<eV}
\left(u_{{\bm k}\uparrow}(x)v^{\ast}_{\bm k\uparrow}(x)-u_{{\bm k}\downarrow}(x)v_{{\bm k}\downarrow}^{\ast}(x)\right)\sin\left(\epsilon_{\bm k}t\right). \label{indf2}
\end{align}
\end{subequations}
Apparently, the bias-dependence of Eqs.~(\ref{pa}) is entirely given
by Eqs.~(\ref{findV}).
\subsection{Spin transport}
\label{spintrans}

We now discuss the appropriate expressions for spin transport quantities.
We expect that with either an 
external bias or a macroscopic phase difference $\Delta\varphi$ between two S banks,
there will be a leakage of magnetism due to a spin-transfer torque~\cite{wvh14,hvw15}.
The local magnetization is related to the spin density and defined as,
\begin{align} \label{mag}
{\bm m}({\bm r})  =-\mu_B\, \langle {\bm \eta}({\bm r})   \rangle\equiv -\mu_B\sum_{ss^\prime}\langle\psi_s^\dagger({\bm r})  {\bsigma}_{ss^\prime} \psi_{s^\prime}({\bm r})\rangle,
\end{align}
where ${\bm \eta}({\bm r}) $ is the spin density operator and $\mu_B$  the Bohr magneton. 
%Since the $yz$ plane is infinite, 
%any spatial variations in the local magnetization
%occurs in the $x$ direction only. 
Again, by using the generalized Bogoliubov transformation, 
each component of ${\bm m}$ can be written 
in terms of the quasiparticle and quasihole wavefunctions:
\begin{subequations}
\label{mmcomp}
\begin{align}
m_x=&- 2\mu_B \sum_n {\rm Re}\left[ u_{n\uparrow} u_{n\downarrow}^\ast f_n -v_{n\uparrow} v_{n\downarrow}^\ast(1-f_n) \right] \\
m_y=& 2\mu_B \sum_n {\rm Im}\left[ u_{n\uparrow} u_{n\downarrow}^\ast f_n +v_{n\uparrow} v_{n\downarrow}^\ast(1-f_n) \right] \\
m_z=&- \mu_B \sum_n \left[ \left(\left|u_{n\uparrow}\right|^2- \left|u_{n\downarrow}\right|^2\right) f_n +\left(\left|v_{n\uparrow}\right|^2- \left|v_{n\downarrow}\right|^2\right)(1-f_n) \right],
\end{align}
\end{subequations}
where we have suppressed the $x$ dependence. 

Using the Heisenberg equation can give the proper
conservation law~\cite{brou,wvh14} for spin densities:
\begin{align} \label{scom}
\frac{\partial}{\partial t} \langle {\bm \eta}({\bm r},t) \rangle = i \langle 
[{\cal H},{\bm \eta}({\bm r},t)] \rangle.
\end{align}
After carrying out some lengthy algebra, we  obtain the desired continuity equation,
\begin{align} \label{scon}
\frac{\partial}{\partial t} \langle {\bm \eta}({\bm r},t) \rangle + \frac{\partial {\bm S}}{\partial x} &= 
{\bm \tau},
\end{align}
where ${\bm S}$ is the spin current and ${\bm \tau}$ is
the associated spin-transfer torque. They are given by
\begin{align} \label{scur}
{\bm S}=\frac{i\mu_B}{2m}\sum_s\left\langle \psi_s^{\dagger}{\bm \sigma}\frac{\partial \psi_s}{\partial x}
-\frac{\partial \psi_s^{\dagger}}{\partial x}{\bm \sigma}\psi_s\right\rangle, %ct	
\end{align} 
\begin{align} \label{stt}
{\bm \tau}
%=-i \langle \psi^\dagger({\bm r}) [{\bm h}\cdot{\bm \sigma},{\bm \sigma}] \psi ({\bm r})\rangle 
=2 \sum_{ss^\prime}\langle \psi_s^\dagger({\bm r}) \left({\bm \sigma}\times{\bm h}\right)_{ss^\prime} \psi_{s^\prime} ({\bm r})\rangle
=2 {\bm m}\times{\bm h}.
\end{align}
The spin current density 
is reduced from a tensor  to a vector  due to the  
quasi-one-dimensional nature of our geometry.
Therefore, the three components of the spin current vector
are associated with those of spin densities and spin current
flowing along the $x$ direction, which is perpendicular to the interfaces.
These three components can also be expressed in terms of 
the quasiparticle and quasihole amplitudes:
\begin{subequations}
\label{scurcomp}
\begin{align}
S_x =& \frac{\mu_B}{2m}\sum_n {\rm Im} \left[ \left(u_{n\uparrow}^* \frac{\partial u_{n \downarrow}}{\partial x}+
u_{n\downarrow}^* \frac{\partial u_{n \uparrow}}{\partial x} \right)f_n\right.  \nonumber \\
&\left.
-\left(v_{n\uparrow} \frac{\partial v^*_{n \downarrow}}{\partial x}+
v_{n\downarrow} \frac{\partial v^*_{n \uparrow}}{\partial x}  \right)(1-f_n) \right], \\
S_y  =& -\frac{\mu_B}{2m}\sum_n {\rm Re} \left[ \left(u_{n\uparrow}^* \frac{\partial u_{n \downarrow}}{\partial x}-
u_{n\downarrow}^* \frac{\partial u_{n \uparrow}}{\partial x} \right)f_n\right.  \nonumber \\
&\left.
-\left(v_{n\uparrow} \frac{\partial v^*_{n \downarrow}}{\partial x}-
v_{n\downarrow} \frac{\partial v^*_{n \uparrow}}{\partial x}  \right)(1-f_n) \right], \\
S_z =& \frac{\mu_B}{2m}\sum_n {\rm Im} \left[ \left(u_{n\uparrow}^* \frac{\partial u_{n \uparrow}}{\partial x}-
u_{n\downarrow}^* \frac{\partial u_{n \downarrow}}{\partial x} \right)f_n\right.  \nonumber \\
&\left.
+\left(v_{n\uparrow} \frac{\partial v^*_{n \uparrow}}{\partial x}-
v_{n\downarrow} \frac{\partial v^*_{n \downarrow}}{\partial x}  \right)(1-f_n) \right].
\end{align}
\end{subequations}
When the junctions are in static equilibrium, the spin-current
does not necessarily vanish because any inhomogeneous
magnetization leads to a non-zero spin-transfer torque
thereby causing a net spin current~\cite{wvh14,hvw15}. 
From Eq.~(\ref{scon}), we  see that ${\bm S}$ 
is a local physical quantity, and $\bm{\tau}$ is
responsible for the change in local magnetizations due
to the flow of spin-polarized currents.
As we shall see in Sec.~\ref{results},
this conservation law (with the source torque term) 
for the spin density is a fundamental relation,
and one has to  ensure that it is not violated when  
studying these transport quantities. 

The above expressions, Eqs.~(\ref{mmcomp}) and Eqs.~(\ref{scurcomp}), %for the spin transport quantities 
are applicable only when the external bias is zero. 
Let us go back and discuss the bias dependence of spin transport quantities for
$F_1F_2S$ tunneling junctions.
As in the discussion on the triplet correlations,
we first define the bias induced magnetization
as $\delta {\bm m}(V)\equiv {\bm m}(V)-{\bm m}_0$,
where ${\bm m}_0$ is given by Eqs.~(\ref{mmcomp})
and ${\bm m}(V)$ is the total magnetization with the presence
of a finite bias.
%when the bias is zero.
%We need here the continuity equation for the local magnetization 
%$\mathbf{m}\equiv-\mu_B\sum_{\sigma}\left\langle\psi^{\dagger}_{\sigma} 
%\bm{\sigma}\psi_{\sigma}\right\rangle$, where $\mu_B$ is the Bohr magneton.
%By using the Heisenberg equation 
%$\frac{\partial}{\partial t}\left\langle\mathbf{m}({\mathbf r})\right\rangle
%=i\left\langle\left[{\cal H}_{eff},\mathbf{m}({\mathbf r})\right]\right\rangle$,
%we obtain the relation: 
%\begin{equation}
%\label{spinconserve}
%\frac{\partial}{\partial t}\langle m_i \rangle+ \frac{\partial}{\partial y} S_i= \tau_i,
%\enspace\enspace i=x,y,z,
%\end{equation}
%where  $\bm{\tau}$ is the spin-transfer torque,
%$\bm{\tau}\equiv 2\mathbf{m}\times\mathbf{h}$, 
% and
%the spin current density $S_i$ is given by,
%\begin{equation}
%S_i\equiv\frac{i\mu_B}{2m}\sum_\sigma\left\langle \psi_\sigma^{\dagger}\sigma_i\frac{\partial \psi_\sigma}{\partial y}
%-\frac{\partial \psi_\sigma^{\dagger}}{\partial y}\sigma_i\psi_\sigma\right\rangle.
%\end{equation}
In the low-$T$ limit, the bias induced magnetization
%$m_i\equiv\sum_{\epsilon_\mathbf{k}<eV}\sum_{\sigma}-\mu_B\langle\psi_\sigma^\dagger\sigma_i\psi_\sigma\rangle_\mathbf{k}$ %kh need to define the subscript k
reads,
\begin{subequations}
\label{mag} 
\begin{align}
\delta m_x=&-\mu_B
\sum_{\epsilon_{\bm k}<eV}\left(u_{{\bm k}\uparrow}^{\ast}u_{{\bm k}\downarrow}
+v_{{\bm k}\uparrow}v_{{\bm k}\downarrow}^{\ast}
+u_{{\bm k}\downarrow}^{\ast}u_{{\bm k}\uparrow}
+v_{{\bm k}\downarrow}v_{{\bm k}\uparrow}^{\ast}\right),\\
\delta m_y=&-i\mu_B 
\sum_{\epsilon_{\bm k}<eV}\left(u_{{\bm k}\uparrow}^{\ast}u_{{\bm k}\downarrow}
+v_{{\bm k}\uparrow}v_{{\bm k}\downarrow}^{\ast}
-u_{{\bm k}\downarrow}^{\ast}u_{{\bm k}\uparrow}
-v_{{\bm k}\downarrow}v_{{\bm k}\uparrow}^{\ast}\right),\\
\delta m_z=&-\mu_B
\sum_{\epsilon_{\bm k}<eV}\left(|u_{{\bm k}\uparrow}|^2
-|v_{{\bm k}\uparrow}|^2
-|u_{{\bm k}\downarrow}|^2
+|v_{{\bm k}\downarrow}|^2\right).
\end{align}
\end{subequations}
%where the first sum in the expressions for $m_i$ denote the ground state
%local magnetization. The second sum appears
%as a consequence of the finite bias between electrodes.
Similarly, we can define the corresponding bias induced spin currents,
$\delta {\bm S}(V)\equiv {\bm S}(V)-{\bm S}_0$,
where ${\bm S}_0$ is identitcal to Eqs.~(\ref{scurcomp}).
The bias induced spin currents are given by
%\begin{equation}
%S_i\equiv\frac{i\mu_B}{2m}\sum_{\epsilon_\mathbf{k}<eV}\sum_{\sigma}\left\langle \psi_\sigma^{\dagger}\sigma_i\frac{\partial \psi_\sigma}{\partial y}
%-\frac{\partial \psi_\sigma^{\dagger}}{\partial y}\sigma_i\psi_\sigma\right\rangle_\mathbf{k},
%\end{equation}
%becomes
\begin{subequations}
\label{spincur}
\begin{align}
\delta S_x&=-\frac{\mu_B}{m}{\rm Im}\left[
\sum_{\epsilon_{\bm k}<eV}\left(u_{{\bm k}\uparrow}^{\ast}\frac{\partial u_{{\bm k}\downarrow}}{\partial y}
+v_{{\bm k}\uparrow}\frac{\partial v_{{\bm k}\downarrow}^{\ast}}{\partial y}
+u_{{\bm k}\downarrow}^{\ast}\frac{\partial u_{{\bm k}\uparrow}}{\partial y}
+v_{{\bm k}\downarrow}\frac{\partial v_{{\bm k}\uparrow}^{\ast}}{\partial y}\right)\right],\\
\delta S_y& = \frac{\mu_B}{m}{\rm Re}\left[
\sum_{\epsilon_{\bm k}<eV}\left(u_{{\bm k}\uparrow}^{\ast}\frac{\partial u_{{\bm k}\downarrow}}{\partial y}
+v_{{\bm k}\uparrow}\frac{\partial v_{{\bm k}\downarrow}^{\ast}}{\partial y}
-u_{{\bm k}\downarrow}^{\ast}\frac{\partial u_{{\bm k}\uparrow}}{\partial y}
-v_{{\bm k}\downarrow}\frac{\partial v_{{\bm k}\uparrow}^{\ast}}{\partial y}\right)\right],\\
\delta S_z& = -\frac{\mu_B}{m}{\rm Im}\left[
\sum_{\epsilon_{\bm k}<eV}\left(u_{{\bm k}\uparrow}^{\ast}\frac{\partial u_{{\bm k}\uparrow}}{\partial y}
-v_{{\bm k}\uparrow}\frac{\partial v_{{\bm k}\uparrow}^{\ast}}{\partial y}
-u_{{\bm k}\downarrow}^{\ast}\frac{\partial u_{{\bm k}\downarrow}}{\partial y}
+v_{{\bm k}\downarrow}\frac{\partial
v_{{\bm k}\downarrow}^{\ast}}{\partial y}\right)\right]. 
\end{align}
\end{subequations}
%The first summations in Eqs.~(\ref{spincur}) represent 
%the static spin current densities when there is no bias.
%The static spin current does not need to vanish, since
%a static spin-transfer torque may exist near the
%boundary of two magnets with misaligned exchange fields.
In short, the finite bias leads to 
a nonequilibrium quasiparticle
distribution for the system, and results in 
non-static spin current densities that are represented
by Eqs.~(\ref{spincur}).
Finally, we note that the spin-transfer torque has to vanish
in the superconductor where the exchange field is
zero. 
%When presenting results below, %kh normalizations introduced below
%we normalize $\mathbf{m}$ by %\cite{Halterman2007}
%$-\mu_B(N_\uparrow+N_\downarrow)$, where the number densities, $N_\sigma$,
%are written,
%$N_\uparrow=k_{F}^3(1+h_m)^{3/2}/(6\pi^2)$ and 
%$N_\downarrow=k_{F}^3(1-h_m)^{3/2}/(6\pi^2)$. 
%Following this convention, we normalize $\bm{\tau}$ by
%$-\mu_B(N_\uparrow+N_\downarrow)E_{F}$ and $\mathbf{S}$
%by $-\mu_B(N_\uparrow+N_\downarrow)E_{F}/k_{F}$.

\section{Results}
\label{results}
\subsection{Tunneling Junctions}
\label{tj}

We begin this section by first discussing our numerical results 
on $F_1F_2S$ tunneling junctions as illustrated in Fig.~\ref{ffs_struct}. 
The thicknesses of $F_1$, $F_2$, and $S$ layers
are taken to be $300/k_F$, $10/k_F$,
and $130/k_F$, respectively. 
These thicknesses are fixed throughout this subsection.
The superconducting coherence length is also fixed to be $100/k_F$.
We consider clean interfaces between these layers. 
In other words,  interfacial scattering
events are not taken into account in this subsection
(the main consequence from these events would be to reduce
the proximity effects). 
For our half-metallic tunneling junctions,
the exchange fields in $F_1$, the layer
that is farthest from the superconductor, is $h_1=E_F$ (see Fig.~\ref{ffs_struct}). 
All energy scales are measured with respect to the Fermi energy. 
As will be demonstrated below,
the spin-valve effect is maximized when the exchange field of
the ferromagnet
$F_2$ is relatively weaker, approximately on the order 
of $h_2=10^{-1} E_F$. 

We are mainly interested in spin transport quantities including magnetization, spin current,
and spin transfer torque. As clearly explained in Ref.~\onlinecite{wvh14}, 
even in the static limit where
the bias across the junction is absent, 
the spin current and the spin transfer torque in
general do not vanish near the interface between two $F$ layers as long as
the magnetic configuration is noncollinear. 
Since dynamical transport properties 
are the main concern in the current work, and in order to clearly see
the bias dependence of these spin-dependent quantities, for most of our results in
this subsection we
will restrict ourselves to the dynamic part that is induced by the external bias. 
For example, the ``induced'' 
magnetizations, $\delta {\bm m}(V)$ are defined in Eqs.~(\ref{mag}).
%as $\delta {\bf m}\equiv {\bf m}(V)-{\bf m}(V=0)$.
We conveniently normalize the magnetization by 
$-\mu_B n_e$, where $n_e=k_F^3/3\pi^2$ is the
%$-\mu_B\left(N_{\uparrow}+N_{\downarrow}\right)$,
%where $N_{\uparrow}=k_F^3\left(1+h_1\right)^{3/2}/6\pi^2$ and 
%$N_{\uparrow}=k_F^3\left(1+h_1\right)^{3/2}/6\pi^2$ are
electron number density.
Similarly, the
induced spin currents, $\delta {\bm S}(V)$,
and the induced STT, $\delta \bm{\tau}\equiv \bm{\tau}(V)-\bm{\tau}(V=0)$,
are normalized by 
%$-\mu_B\left( N_{\uparrow}+N_{\downarrow} \right) E_F/k_F$
%and by $-\mu_B\left( N_{\uparrow}+N_{\downarrow} \right) E_F$, respectively.
$-\mu_B n_e E_F/k_F$,
and by $-\mu_B n_e E_F$, respectively.
Below we shall discuss the position dependence of all spin transport quantities.
For convenience, we measure lengths in units of $k_F^{-1}$ and
use $X\equiv k_Fx$ to denote positions.

\begin{figure*} %kh dashed vertical lines identifying each region would be helpful
\includegraphics
[width=5.5in]
{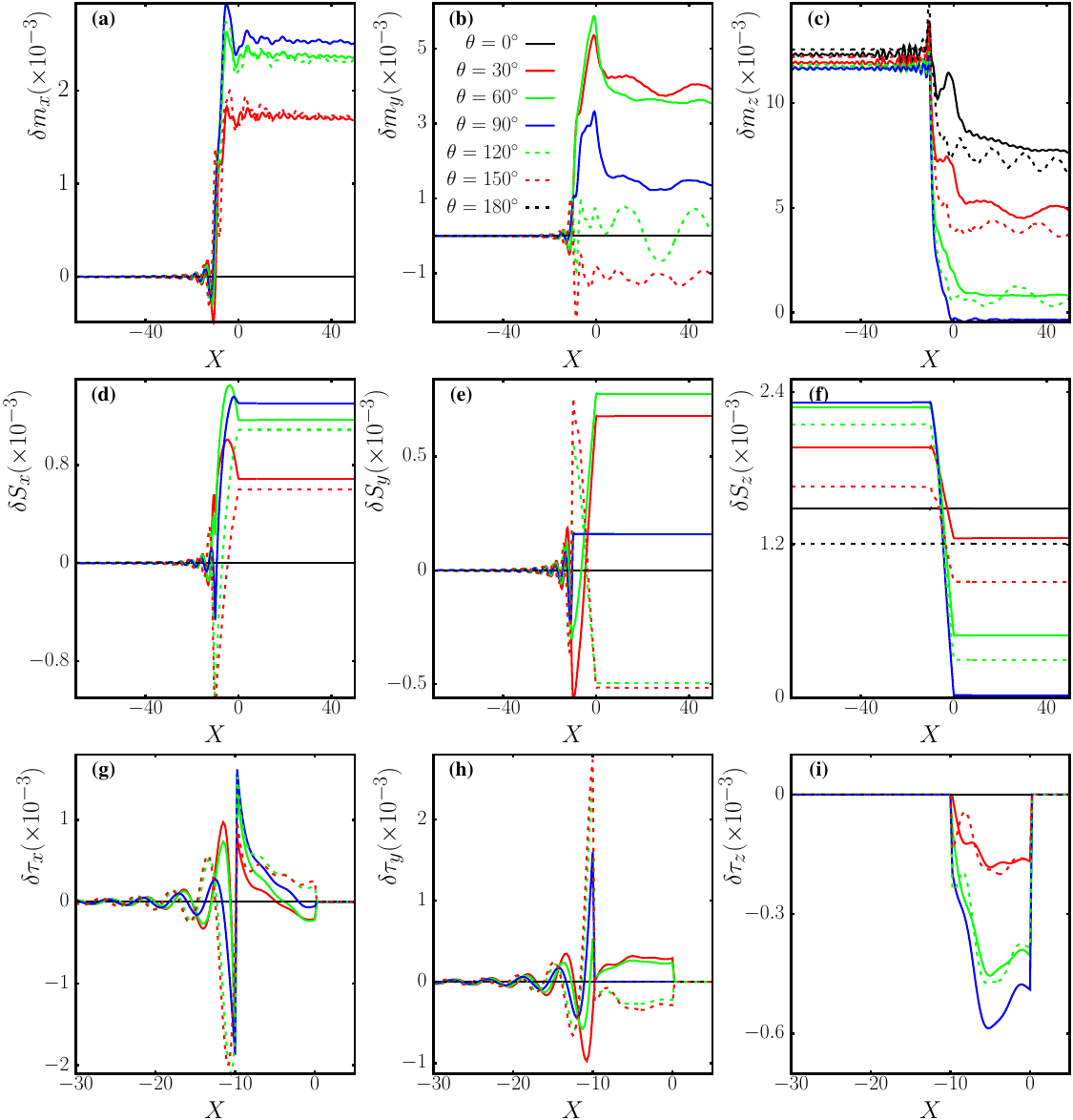}
\caption{
(Color online) 
In this figure, we present spin transport quantities
as functions of position, $k_Fx\equiv X$, 
for several relative angles, $\theta$, between the exchange fields %kh
in the $F_1$ and $F_2$ layers of half-metallic $F_1F_2S$
tunneling junctions. 
The external bias is set to be twice of the bulk superconducting
pair amplitude, $V=2\Delta_0$. The thicknesses of 
${F_1}$, ${F_2}$, and $S$ are set to be
$300/k_F$, $10/k_F$, and $130/k_F$, respectively.
Panels (a)-(c) in the first row  %kh
show the dynamical part, $\delta {\bm m}$, 
of the three magnetization components, computed from  %kh
Eqs.~(\ref{mag}).
Panels (d)-(f) in the second row  %kh
depicts each  component of the dynamical part of the spin currents, %kh
$\delta {\bm S}$, according to Eqs.~(\ref{spincur}).
Spin currents are in general third-rank tensors in three-dimensional 
space. However, since our system is quasi-one-dimensional,
they are reduced to three-dimensional vectors.
Panels (g)-(i) in the third row  %kh
presents the dynamical part of the three components of %kh
the spin-transfer 
torque $\delta {\bm \tau}$, by using
the relation $\delta {\bm \tau}=\delta {\bm m}\times{\bm h}$.
From the figure, one can easily verify the formula  %kh
$\delta \tau_i=\partial S_i/\partial x$
for all $\theta$.
}
\label{fig:1}
\end{figure*}
In Fig.~\ref{fig:1}, we present the angular dependence of the
induced magnetizations,
spin currents, and spin-transfer torques for 
the half-metallic spin valve shown in Fig.~\ref{ffs_struct}.
The half-metallic layer $F_1$ is adjacent to a thinner and relatively weak
ferromagnet with
$h_2=10^{-1}E_F$. 
We begin by giving simple physical reasons
for choosing these parameters. 
The thickness of $F_2$ is chosen to be
thin compared to $F_1$ and $S$ in order to 
take advantage of the superconducting proximity effects. 
For the same reason, the exchange field in $F_2$ also needs to 
be weak enough to study the interplay between the superconducting proximity effects
and spin-valve effects. 
In our coordinate system,
$X=0$ corresponds to the interface between $F_2$ and $S$.
Therefore, in Fig.~\ref{fig:1}, the half-metal $F_1$
lies in the range $X<-10$, the superconductor is in
the region $X>0$, and the $F_2$ layer is in the region
$-10<X<0$. 
The bias across the junction is set to be $2\Delta_0$
in the figure, where $\Delta_0$ is the singlet pair amplitude
in the bulk limit.
Recall that in our considerations, the exchange field in $F_1$ is along
the $\hat{z}$ axis and in $F_2$ it is tilted with respect to the $\hat z$
axis by an angle $\theta$ in the $yz$ plane. %parallel to the interface 
There are two main effects that need to be taken into account 
in order to understand the induced magnetizations:
First, the magnetic moments in $F_1$ and $F_2$ interact,
with the magnetization of $F_1$ leaking  into $F_2$, and vice versa, %kh
resulting in spatial precession. Secondly, both the direction and
magnitude of the static magnetic moments in $F_2$ will 
affect any induced magnetizations when an external bias is present.

For the three components of the
induced magnetizations (Panels (a)-(c) in Fig.~\ref{fig:1}), %ct 
we first see that $\delta m_x$ and $\delta m_y$ 
vanish throughout the entire junction when $\theta=0^{\circ}$ and $180^{\circ}$.
This is because the contributions from both the precession and static magnetizations
are zero when the exchange fields are 
parallel ($\theta=0^{\circ}$)
or anti-parallel ($\theta=180^{\circ}$) to each other. 
Let us first focus on $\delta m_x$ for other relative angles. 
The magnitudes for $\theta$ and $\pi-\theta$ 
are of the same order in the $S$ region because %can be explained purely by the precession effect:
the $x$ component of the static magnetization
is not present (recall that the exchange fields in our system are always in-plane)
and only the precession effect is at work.
Turning to the
$\delta m_y$ panel,
its magnitude in $S$ for $\theta=90^{\circ}$ (the exchange field in $F_2$ is along $y$) is  
determined purely from  the static magnetization
because the precession effect will only affect $\delta m_x$ and $\delta m_z$ at
this angle. 
Physically, this tells us that the system becomes spin-polarized 
in the $xy$ plane in $S$. 
When $90^{\circ}<\theta<180^{\circ}$,
the contribution to $\delta m_y$ from the precession effect is negative
while the contribution from the effect of the static magnetization in
$F_2$ is positive.
The cumulative result is that the magnitudes 
are much smaller than
their counterparts for $0^{\circ}<\theta<90^{\circ}$ in the $S$ region.
For $\delta m_z$, we can see that
it is the only non-zero component throughout the junction for parallel ($\theta=0^{\circ}$)
and anti-parallel ($\theta=180^{\circ}$) configurations.
The behaviors for other relative angles are simply explained
again by the precession effect, just as in the case for $\delta m_x$.

Next, we analyze the behaviors of the
induced spin currents and spin transfer torques.
The spin-transfer torques are determined by the expression, Eq.~(\ref{stt}),
which are in turn
related to the spin currents  given in Eqs.~(\ref{scurcomp}) and~(\ref{spincur}).
This is clearly seen 
in the steady state, where their interplay is encapsulated by the
expression, $\frac{\partial\bm S}{\partial y}=\tau$.
More generally, one can intuitively understand the role of
the induced spin currents $\delta {\bm S}$ by considering
the static magnetizations in each of the ferromagnetic layers.  
The $F_1$ layer is relatively thick, and can be regarded as a spin source, which
polarizes the incoming current along the $+z$ direction. When a spin current originating from 
$F_1$ flows into $F_2$, the polarization state  can be rotated by 
means of the local exchange field in $F_2$ and corresponding induced STT. 
For the $z$ component of the induced spin currents, $\delta S_z$,
at $\theta=0^{\circ}$, it is constant throughout the entire junction
including the superconducting layer as the spin density along $z$
commutes with the Hamiltonian. The same argument 
holds for the other collinear orientation $\theta=180^{\circ}$.
However, the magnitude of $\delta S_z$ is larger at $\theta=0^{\circ}$ 
than at $\theta=180^{\circ}$, as a consequence of the exchange
fields in the $F_1$ and $F_2$ layers being oppositely directed
while $h_1\gg h_2$. 
In fact, the magnitude of $\delta S_z$ is higher when $\theta<90^{\circ}$
than the counterparts at $\pi-\theta$, for exactly the same reasons.
Although $\delta S_z$ at $\theta=90^{\circ}$  %kh
vanishes inside the superconductor, 
we found that in general, this is not necessarily 
the case. The magnitude and the sign of $\delta S_z$ depend on 
both the thickness of $F_2$ and the strength of the 
exchange field.
Thus, by carefully choosing the thickness of the second 
ferromagnet, which plays an important role in both triplet proximity
effects and spin-transfer torques, in principle
the spin transport properties of spintronics devices
can be manipulated experimentally.

Let us now turn our attention to the remaining components, $\delta S_x$ and $\delta S_y$. 
In the collinear configurations ($\theta=0^{\circ}$ and $\theta=180^{\circ}$), 
both the $x$ and $y$ components are zero
because of the absence of the precession effect.
Both the sign and magnitude of $\delta S_y$ in the $S$ region roughly
follow the $y$ component of the exchange field in $F_2$.
Although the $y$ component of the exchange field in $F_2$ 
is at its maximum when $\theta=90^{\circ}$,
we find that the corresponding $\delta S_y$ in $S$ is smaller than when at the other angles.
This is because  when $\theta\neq 90^{\circ}$, the $y$ component
of the spin density can still be induced via the  spin density precession
coming from the half-metallic layer that possesses a
much larger magnetization strength, which in turn 
is more dominant than the other effect.
For the same reason, $\delta S_y$ in $S$ is higher at $\theta$ than at $\pi-\theta$,
where $\theta<90^{\circ}$.
The precession effect is seen to play an important role as well in 
the behavior of $\delta S_x$, where as panel (d) shows,  at $\theta=90^{\circ}$, the dynamical 
part  $\delta S_x$ abruptly increases in $F_2$, and
then uniformly extends into the $S$ region where it is  maximized.

\begin{figure*}
\includegraphics
[width=5.5in]
{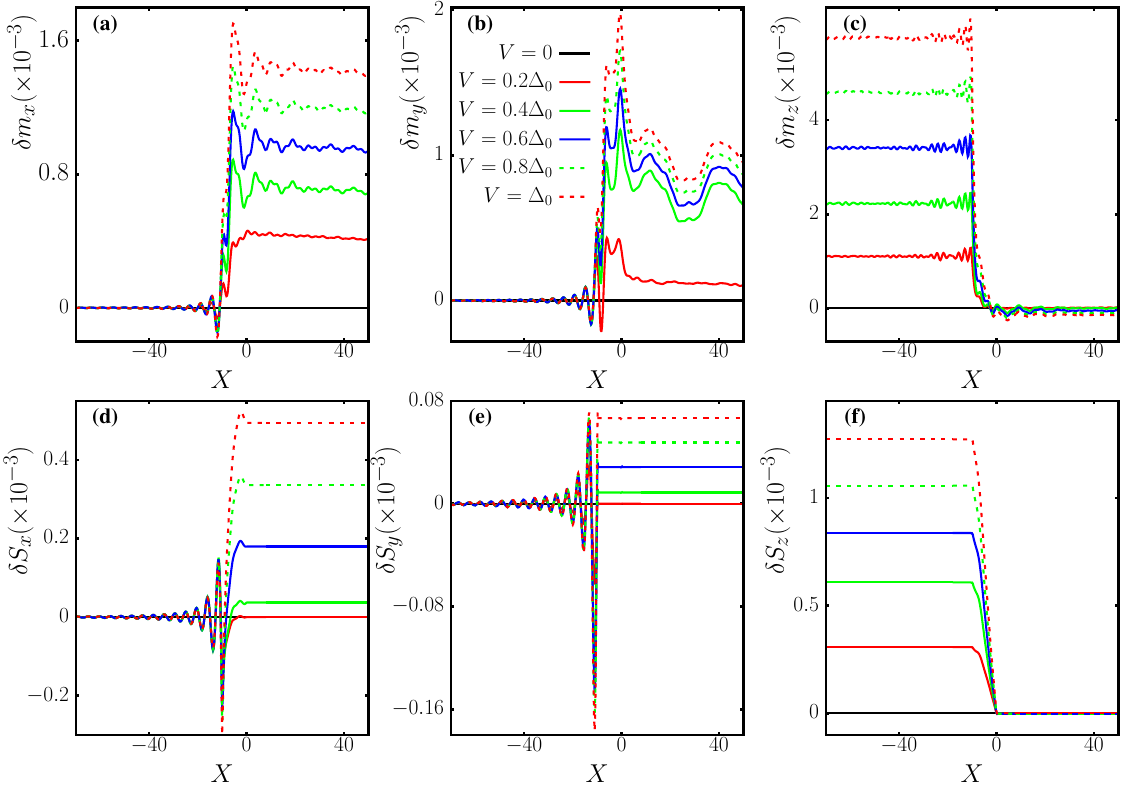}
\caption{
(Color online) 
In this figure, we present spin transport quantities
as functions of position, $k_Fx\equiv X$,  %kh
for several external biases, $V$, 
scaled by the bulk superconducting %kh
gap, $\Delta_0$, in half-metallic $F_1F_2S$ %kh
tunneling junctions. 
The relative angle $\theta$ between exchange fields
in $F_1$ and $F_2$ is set to be $90^{\circ}$. 
The thicknesses of $F_1$, $F_2$, and $S$ are set to be
$300/k_F$, $10/k_F$, and $130/k_F$, respectively.
Panels (a)-(c) in the first row  %kh
show the dynamical part, $\delta {\bm m}(V)$, 
of the three magnetization components, computed from  %kh
Eqs.~(\ref{mag}).
Panels (d)-(f) in the second row 
show the dynamical part, $\delta {\bm S}(V)$, 
of the three spin current components, computed from %kh
Eqs.~(\ref{spincur}).
}
\label{fig:2}
\end{figure*}
The last interesting quantity is the spin-transfer torque,
which is numerically determined using the relations involving  
the self-consistently calculated  $\delta \bm m$ and the exchange field $\bm h$ [see Eq.~(\ref{stt})].
Since $\bm h$ vanishes identically inside the superconductor,
all components of $\delta \bm \tau$ must vanish there. 
The absence of a torque in the superconductor imposes that the
spin current there cannot vary in space as Eq.~(\ref{scon}) shows.
Thus the constancy of the spin currents 
inside the superconducting region shown in Fig.~\ref{fig:1}.
It is also straightforward to understand why $\delta \tau_z=0$ in the half-metal $F_1$.
%when $\theta=0^{\circ}$ or $\theta=180^{\circ}$.
We find that $\delta \tau_z$ is maximized in $F_2$
when $\theta=90^{\circ}$, suggesting
that the corresponding $\delta S_z$ must have the greatest 
change in $F_2$. Indeed, as can be seen in panel (f),
the only spatially varying region is in the ferromagnet $F_2$,
and it occurs the greatest when $\theta=90^{\circ}$.
We emphasize here that the static part of $\tau_x$ is 
in general non-vanishing as long as the in-plane 
exchange fields are non-collinear in the ${F_1F_2S}$
tunneling junctions.
The static part of $\tau_x$ is much larger than
the dynamic part. Therefore, the behavior $S_x$ 
does not significantly change with the presence
of bias (not shown).
In panel (e), it was observed that
 the precessional effect combined with the magnetization rotation in $F_2$,
 led to  a reversal in the bias-induced spin current variation as $\theta$ changed. 
 These  abrupt changes  in $\delta S_x$ translate into torque reversals
within the relatively weaker ferromagnet region, as well as drastic
variations near the $F_1$/$F_2$ interface, as demonstrated in (h).

In the linear-response regime, transport quantities
are in principle dependent on the external bias, $V$.
However, with the presence of superconductors, transport quantities
sometimes exhibit distinct behavior above and below
the superconducting gap.
The related transport phenomena including excess current and tunneling conductance
are thoroughly discussed in Refs.~\onlinecite{wvh14,btk}.
This gap-dependent feature can be attributed to Andreev reflections. 
When the external bias is below the superconducting gap,
current is not suppressed due to the mechanism of the Andreev scattering.
Once the external bias is above the gap, the 
contribution to current from  ordinary scattering emerges.
As explained in Sec.~\ref{methods},
the superconducting pair amplitudes are determined self-consistently
and the gap profiles are position-dependent,  
which saturate deep inside the bulk superconductor. The
saturation values of the gap profiles are important and usually
smaller than the bulk superconducting gap, $\Delta_0$.
Furthermore, the saturation values also depend on the relative magnetization
angle, $\theta$. 

\begin{figure*}
\includegraphics
[width=5.5in]
{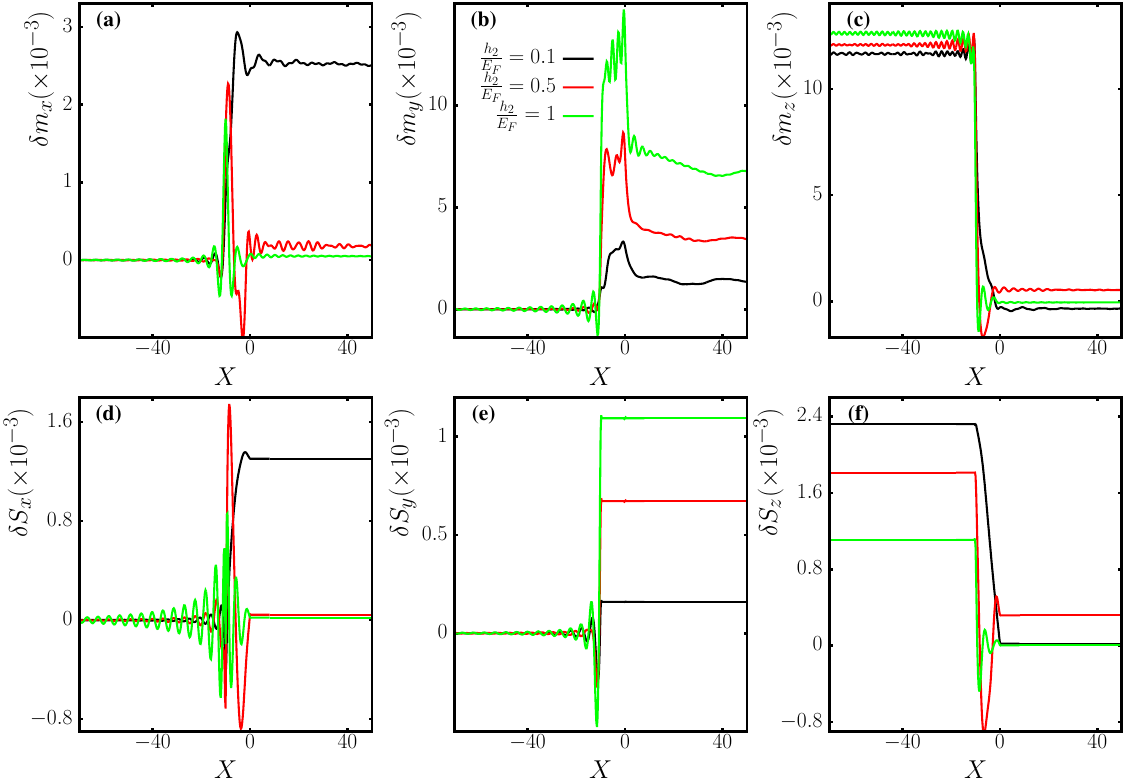}
\caption{
(Color online) 
In this figure, we present spin transport quantities
as functions of position, $k_Fx\equiv X$,  %kh
for three different $h_2$ measured in terms of the Fermi energy
for half-metallic $F_1F_2S$ tunneling junctions. 
The external bias is fixed to be twice  the bulk superconducting
gap, $V=2\Delta_0$.  %kh
The relative angle $\theta$ between exchange fields
in $F_1$ and $F_2$ is also fixed and its value is $90^{\circ}$. 
The thicknesses of ${F_1}$, ${F_2}$, and $S$ are set to be
$300/k_F$, $10/k_F$, and $130/k_F$, respectively.
Panels (a)-(c) in the first row  %kh
show the dynamical part, $\delta {\bm m}(V=2\Delta_0)$, 
of the three magnetization components, computed from 
Eqs.~(\ref{mag}).
Panels (d)-(f) in the second row  %kh
show the dynamical part, $\delta {\bm S}(V=2\Delta_0)$, 
of  the three spin current components, computed from  %kh
Eqs.~(\ref{spincur}).
}
\label{fig:3}
\end{figure*}
In Fig.~\ref{fig:2}, we plot spin transport quantities  at
several different biases for $\theta=90^{\circ}$. The thicknesses
of each layer and exchange interactions are the same as in Fig.~\ref{fig:1}. %ct
Our self-consistent calculations reveal that
the saturation value for the superconducting gap is approximately 
$0.3\Delta_0$. First, we note the trivial fact
that the dynamic part of
all spin transport quantities vanishes when $V=0$. We then pay
particular attention to the behavior above and below the saturation point $0.3\Delta_0$.
Note that all three components of $\delta \bm m$ do not significantly
change qualitatively with increased bias,
and the major quantitative change is their magnitudes.
Nevertheless, $\delta m_y (V=0.2\Delta_0)$ is greatly suppressed compared to 
$\delta m_y (V>0.2\Delta_0)$ while $\delta m_x (V=0.2\Delta_0)$ is not. 
We also see that the magnitudes of both $\delta m_x$ and $\delta m_y$ 
increase linearly with $V$ for $V>0.3\Delta_0$.
On the other hand,
$\delta m_z$ does not show very distinct behavior
above or below $0.3\Delta_0$, and it increases linearly 
in the entire $V$ range we considered here.

For the dynamic part of the
spin currents $\delta {\bm S}$, we find
that $\delta S_x$ and $\delta S_y$ 
disappear 
inside the superconducting
region when $V<0.3\Delta_0$. This is 
due to the fact that any spin polarized current entering the superconductor 
is converted into a supercurrent, which is spin unpolarized.
For $V>0.3\Delta_0$,
the magnitudes inside the superconductor increase linearly with the bias,
similar to what was found for 
$\delta m_x$ and $\delta m_y$.
At these larger bias voltages, 
 $\delta S_x$ and $\delta S_y$ within the half-metal are
insensitive to changes in $V$.
Examining  panel (f),
the current entering the $F_1$ region becomes strongly polarized 
by the half-metal, and  $\delta S_z$ increases nearly linearly 
with greater bias before decaying away after
interacting with the adjacent ferromagnet whose 
exchange field is orthogonal to it (along $y$).
It is evident  that unlike $\delta S_x$, there
are no
abrupt changes
in behavior about
the saturation point 
$0.3\Delta_0$.
Examining the top row of Fig.~\ref{fig:2}, one can infer
the qualitative behavior of the torque throughout the structure.
Thus, the bias dependence to the spin transfer torque 
is omitted here, as it clearly 
follows that of $\delta {\bm m}$. 
%It is not unexpected
%because the spin transfer torque is determined by $\delta {\bf m}$.

Next, we explore spin transport properties 
with different strengths of the exchange field in $F_2$ while fixing the exchange
field in $F_1$ to be $h_1=E_F$. 
In Fig.~\ref{fig:3}, we plot $\delta {\bm m}$ (top row) and $\delta {\bm S}$ (bottom row)
for three different $h_2$. The relative angle between 
the  exchange fields in $F_1$  and $F_2$
is again fixed at  $\theta=90^{\circ}$ (the direction of the exchange interaction in $F_2$ is along $y$), and %ct %kh2
the bias is set at $V=2\Delta_0$.
In panel (b), 
we see that the overall trends in the induced magnetization do not change
significantly 
for different $h_2$,
where $\delta m_y$ is damped out in the half-metal, and then peaks in 
 $F_2$ before propagating into the superconductor.
The half-metal has its exchange field aligned in the $z$ direction, thus
the current is initially polarized in this direction leading to a nearly vanishing 
$y$ component of the induced magnetization, which becomes $y$ polarized when entering 
adjacent ferromagnet.
The result is that  $\delta m_y$ from both $F_1$ (due to the precession effect)
 and $F_2$ (due to the inherent magnetization)
extend  into the superconductor with a magnitude
 proportional to $h_2$.
For  
the induced magnetization normal to the interfaces, $\delta m_x$,  
we see that it builds up within $F_2$,
and then undergoes  damped oscillations (see panel (a)).
The 
period of these oscillations in $F_2$
are governed by the degree of spin polarization
in the ferromagnet and thus
scale inversely proportional
to $h_2$. Therefore, one can see that for such 
a thin  $F_2$, $\delta m_x$ with
$h_2=0.1$ is too confined to  possess even a full period of oscillation.
As a result, when $h_2=0.1E_F$, $\delta m_x$ becomes ``squeezed"
and has a  larger 
magnitude in $F_2$
compared to when $h_2=0.5E_F$ and $h_2=E_F$.
If we increase the thickness of $F_2$,
$\delta m_x$ for $h_2=0.1E_F$ will also 
become negligible inside the $S$ layer.
This property provides a way for experimentalists
to control the flow of magnetization by 
varying the thickness of the intermediate ferromagnetic
layer. 
Turning now to panel (c), it is seen that
inside $F_1$, $\delta m_z$ is only very weakly
dependent on $h_2$ and  is uniform in space.
Inside $F_2$ it exhibits damped oscillations,
akin to $\delta m_x$,  with an oscillation period
that is inversely proportional
to $h_2$. If the $F_2$ layer is thick enough,
$\delta m_z$ will vanish identically inside 
the $S$ layer, irrespective of  $h_2$.
This sensitivity to thickness can be used to control not only whether 
$\delta m_z$  vanishes in the $S$ layer, but also %ct
for appropriate 
 thicknesses, 
whether it can be positive or negative.

Now, let us compare spin currents for different $h_2$.
From panel (e), we see that for a given $h_2$,
the induced
$\delta S_y$ is constant and flows uninterrupted 
 inside both the $F_2$ and $S$ layers.
 This is a  reflection of  the fact
that the $y$ component of the spin-transfer torque vanishes in those regions.
As Eq.~(\ref{stt}) showed, this can also be found by simply computing 
the cross product between $\delta \bm m$ and $\bm h$.
For the same reasons, $\delta S_x$ is constant inside
the $S$ layers only, while $\delta S_z$ is constant
in the $F_1$ and $S$ regions.
For each $h_2$, the relative magnitudes of $\delta \bm S$ 
in $F_2$ and the superconducting region follow  similar trends
as $\delta \bm m$, in that 
there is a positive correlation between 
the corresponding components of  $\delta \bm S$ and $\delta \bm m$.
We also find that the spatial period for the oscillation
inside the $F_2$ layer is the same as that of $\delta \bm m$
for a given $h_2$. Finally,  it is important to stress  that
both the direction and the magnitude of $\delta \bm S$
can also be adjusted by changing the $F_2$ thickness.
In practice, one would like to choose a weaker ferromagnet
for this intermediate layer. 
This follows not only from the potential triplet pair enhancement (discussed below), 
but also when
a strong ferromagnet
is adopted, the $F_2$ thickness should be relatively thin
in order to take advantage of this thickness sensitivity.
As before, we do not present the spin-transfer torques here since
they can be computed directly from  knowledge of $\delta \bm m$ [Fig.~\ref{fig:3}, first row],
and $\bm h$.

We now focus on the induced triplet correlations for these
half-metallic tunneling $F_1F_2S$ junctions. 
It is useful to recall that 
the triplet correlations can be induced
even 
in the absence of an external bias~\cite{wvh12}.
As discussed in Ref.~\onlinecite{wvh12},
triplet correlations with $m=\pm 1$  projections  
on the spin quantization axis are important
since these spin-polarized pairs  are immune to pair-breaking effects
of the exchange fields in the $F$ layers. This is especially relevant when
a very strong half-metallic layer is present. 
Successful control of a dissipationless supercurrent
is regarded as one of the essential goals 
in the development of practical  low-temperature  spintronics devices. 
Presumably, this can be achieved by generating and controlling  
the $f_1$ and $f_2$ equal-spin triplet pairs,~\cite{hvw15} 
since they are able to propagate over relatively long distances
without serious degradation.
To simplify the discussions below,
we  shall focus on the  $f_1$ equal-spin and $f_0$ opposite-spin triplet channels, since in many cases
$f_2$ behaves complimentary to $f_1$.

The physics of induced triplet correlations for spin valves
in the static limit has been extensively discussed in Ref.~\onlinecite{wvh12}.
Also, we find that
in the $F_1$ layer the dynamic part  
is added constructively
to the static part of the triplet amplitudes.
Therefore, we focus here on the dynamical situation where the external
bias is non-vanishing and confine our attention to
the dynamic part of the induced triplet correlations.
To find the bias dependence to the
triplet pairs in our system,
we define, similar to previous quantities, the induced 
triplet correlations via
 $\delta f_i(V)=f_i(V)-f_i(V=0)$, where 
$i=0,1$.

\begin{figure}
\includegraphics
{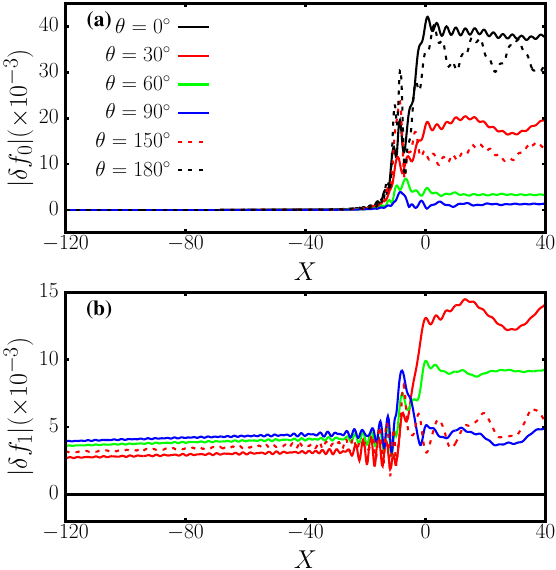}
\caption{
(Color online) 
The dynamical part of the
induced triplet correlations 
as functions of position, $X$, 
for several angles $\theta$.
In panel (a) we have $\delta f_0(V=2\Delta_0)$ [see Eq.~(\ref{indf0})],
and
in panel (b) we have  $\delta f_1(V=2\Delta_0)$ [see Eq.~(\ref{indf1})]. 
% between exchange fields in
%$\rm F_1$ and $\rm F_2$ for half-metallic $\rm F_1F_2S$ tunneling junctions. 
The external bias is fixed to be twice that of the bulk superconducting gap, $V=2\Delta_0$. 
The relative time of these triplet correlation is $\omega_D t=4$.
The thicknesses of $F_1$, $F_2$, and $S$ are set to be
$300/k_F$, $10/k_F$, and $130/k_F$, respectively.
The exchange fields are $h_1=E_F$ and $h_2=0.1E_F$.
}
\label{fig:4}
\end{figure}
In Fig.~\ref{fig:4}, 
we present the angular dependence of 
both the opposite-spin $f_0$ and
equal-spin $f_1$  
triplet pairs. 
The pair correlations are functions of their relative  time difference $t$, 
which is
set according to the dimensionless relation $\omega_D t=4.0$.
The external
bias is fixed at $V=2.0\Delta_0$. 
The thicknesses are the same as in previous figures, with
the exchange fields in $F_1$ and $F_2$ again corresponding to $h_1/E_F=1$ and $h_2/E_F=0.1$,
respectively.
For  $\delta f_0$ shown in the top panel (a),
we find that it decays into the half-metallic layer
with a very short decay length, as it is energetically unstable
due to the presence of a single spin band at the Fermi level.
Within the thin ferromagnet ($-10<X<0$), $\delta f_0$ is largest when 
 a single quantization  axis can be ascribed to the system,
 i.e., 
 when
the magnetizations of both $F$ layers are collinear.
There is a
 slightly
more  pronounced effect when 
$\theta$ corresponds to the antiparallel  configuration, where there
are greater competing effects between 
the magnetizations in the  $F_1$ and $F_2$  layers.
When $F_1$ and $F_2$ are in
the orthogonal configuration with $\theta=90^\circ$,
they are then in 
their most inhomogenous magnetic state, and 
the $\delta f_0$ amplitude is lowest  in $F_2$.
For other orientations that are closest to the orthogonal configurations, 
such as $\theta=60^\circ$, $\delta f_0$ is also relatively weak compared 
to the collinear situation, but larger compared to $\theta=90^\circ$ due  
a finite  $z$ component to the magnetization.
These findings for the thin ferromagnet  layer
carry over to 
the superconducting layer,
where 
the following  angular dependence is observed: $\delta f_0$ is minimized
at $\theta=90^{\circ}$ (orthogonal configuration)
and maximized for the collinear configurations 
($\theta=0^{\circ}$ and $180^{\circ}$). 

We turn now to the
more interesting $\delta f_1$ component,  which is much more robust
against the magnetic pair-breaking effects. In the bottom panel (b), 
we present  the spatial behavior of $\delta f_1$, again for several $\theta$.
We first see that $\delta f_1$ vanishes for the collinear configuration,
as it should, as
 explained earlier in the introduction. 
For other relative angles, $\delta f_1$ is generated  because of
the non-collinear magnetic profile which prevents
the system from being described by a single quantization axis. 
Furthermore, as shown in panel (b),
the bias-induced $\delta f_1$ triplet amplitude 
is long-ranged in the half-metal and maximized 
for orientations around $\theta=90^{\circ}$.
This is the central result of this subsection.
Once the spin-polarized triplet pairs pass through $F_2$,
they enter the superconductor
and become enhanced, not for the orthogonal configuration, 
but rather for slight misalignments in the relative magnetizations.
These trends are similar to what was observed in Fig.~\ref{fig:1} for
the $y$-component of the bias induced magnetization.
Thus, we have demonstrated the long-range nature of the
dynamic part of the triplet pairs by showing that only the $\delta f_1$ component survives in
the half-metal. Also, due to the interactions between $F$ layers and triplet conversion effects,
spin-polarized triplets were shown to be effectively generated within the $S$ region. 
Moreover, our study revealed 
that $\delta f_0$ in $F_2$ and $S$,  and $\delta f_1$ in the half-metal are often anticorrelated,
i.e.,when $\delta f_0$ is maximized (minimized), $\delta f_1$ 
is minimized (maximized).  

\begin{figure}
\includegraphics
{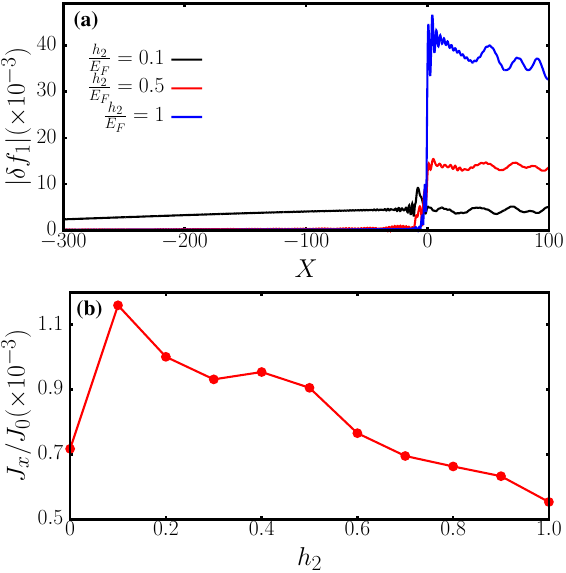}
\caption{
(Color online) 
In  panel (a) of this figure, we present the dynamical part of %kh
induced triplet correlations, $\delta f_1$,
as functions of position, $X$, %kh %ct
for three different normalized exchange fields in the $F_2$ layer: %kh %kh2
$h_2/E_F=0.1$, $h_2/E_F=0.5$, and $h_2/E_F=1.0$. %kh2
%in half-metallic $F_1F_2S$ tunneling junctions.
The angle $\theta$ between exchange fields in
$F_1$ and $F_2$ is chosen to be $\theta=90^{\circ}$.
The external bias is fixed at twice the bulk gap, %kh
$V=2\Delta_0$. 
%The pairing time of these triplet correlation is $\omega_D t=4.$ %kh
The thicknesses of $F_1$, $F_2$, and $S$ are set at
$300/k_F$, $10/k_F$, and $130/k_F$, respectively.
In panel (b), we show the normalized charge current density v.s. the exchange field %kh
$h_2$ of the intermediate $F$ layer.
The data points are connected by lines
to 
serve as guides to the eye.
The relative angle between the exchange fields in
$F_1$ and $F_2$ is $90^{\circ}$.
%The thicknesses of $F_1$, $F_2$, and $S$ are set to be %kh
%$300/k_F$, $10/k_F$, and $130/k_F$, respectively.
}
\label{fig:5}
\end{figure}
It was mentioned at the beginning of this subsection that
our choice of $h_2/E_F=0.1$ for the exchange field strength in the thin intermediate $F_2$ layer, 
resulted in the optimal amount of spin-polarized pairs in
the half-metallic region. 
To illustrate this,
it is insightful to consider differing exchange field magnitudes
in $F_2$ and  examine how these differences affect
the equal-spin triplet pairs throughout the entire junction.
Thus,
we present  in panel (a) of Fig.~\ref{fig:5}, the spatial dependence of the
magnitude of the
dynamic part $\delta f_1$ for 
several $h_2$. 
We set
$\theta=90^{\circ}$, 
creating the most magnetically 
inhomogenous configuration possible,
and thus maximizing
$\delta f_1$ in $F_1$.
Note that here  
the spatial range is much 
wider than the results
presented before in order 
to identify  
any long range 
behavior  of 
the
spin-polarized
triplet correlations.
First, inside the superconducting layer, 
we find that
the magnitude of $\delta f_1$
is approximately  proportional to $h_2$.
However, in the non-superconducting regions, $\delta f_1$ for
both $h_2/E_F=0.5$ and $h_2/E_F=1.0$ decays with a very small
characteristic decay length. On the other hand, the weaker
exchange field of $h_2/E_F=0.1$ results in
$\delta f_1$ penetrating quite extensively into the $F$ regions,
thereby establishing its long range behavior.
This result is significant, and it justifies our choice of for $h_2$,
mentioned earlier. Although we do not show
the static part of the induced triplet correlations,
we find the same behavior as before: the static part of $f_1$
is long-ranged when the magnetic configuration is 
non-collinear and its magnitude is comparable
to the dynamic part.
% is about the same order
%of magnitude as the dynamic part.
In the absence of a bias voltage, the 
corresponding static $f_1$ amplitudes 
are also maximized  when $h_2\sim0.1E_F$. %kh correct?

%\begin{figure} %kh it would be interesting to plot the z-comp. of spin current in the half metal as a function of h_2
%kh, or the x-component of spin in the superconductor vs h_2
%kh and see how it compares to the charge current jx
%\includegraphics
%[width=3.2in]
%{fig6.pdf}
%\caption{
%(Color online)
%The normalized charge current density v.s. the exchange field 
%$h_2$ for the intermediate F layer.
%The data points are connected by lines
%to 
%serve as guides to the eye.
%The relative angle between the exchange fields in
%$F_1$ and $F_2$ is $90^{\circ}$.
%The thicknesses of ${\rm F_1}$, ${\rm F_2}$, and ${\rm S}$ are set to be
%$300/k_F$, $10/k_F$, and $130/k_F$, respectively.
%}
%\label{fig:6}
%\end{figure}
To further corroborate these  ideas, we show
in panel (b) of Fig.~\ref{fig:5}
the  charge current density, $J_x$, along the direction 
perpendicular to the interface as a function of
$h_2$. 
The current density is normalized by
$J_0\equiv e n_e v_F$, where $n_e$ is the electron density %ct
and $v_F\equiv k_F/m$ is the Fermi velocity.
Here we fix the external bias
to be $V=2.0\Delta_0$ and the relative angle between
the exchange fields in the $F_1$ and $F_2$ layers
is $\theta=90^{\circ}$.
As in Refs.~\onlinecite{hvw15} and \onlinecite{wvh14},
it is stressed that the current density is spatially
uniform throughout the junction in order to satisfy
the continuity equation. 
In the $S$ region
one should consider both
the current density computed from Eq.~(\ref{current2})
and also the integration of the source term in
Eq.~(\ref{current}), since the pair potential is not zero there.
To avoid this complexity, we compute the current density
from Eq.~(\ref{current2}) directly in the $F$ region. 
Furthermore, we verify that if one includes the contribution
from the source term, the current density is indeed uniform
across the entire tunneling junctions.
From panel (b) of Fig.~\ref{fig:5}, we find that the current density
is maximized at $h_2/E_F=0.1$. Recalling that the equal-spin triplet correlations $f_1$
are the
most long-ranged at $h_2/E_F=0.1$, this suggests a correlation
between the long-ranged nature of the spin-polarized triplet pairs and
the charge transport.
Finally, we see that the charge density is lowest at $h_2/E_F=1$,
where only one spin band is accessible in both $F$ layers for the  current carrying  states.
The results presented in Fig.~\ref{fig:5} %and \ref{fig:6} 
therefore 
strongly suggest that by using relatively thin ferromagnets with  weak
exchange fields,
the half-metallic region will effectively host long-range spin-polarized triplet pairs
that offer hints of their signatures in the charge transport behavior.
Thus, to achieve these properties for the structures considered here, $h_2/E_F=0.1$ is the optimal strength for such half-metallic
superconducting spintronic devices.
If on the other hand it is desired to generate $f_1$ triplet pairs solely 
in the superconductor, one should incorporate half-metals into both $F$ regions.

\subsection{Half Metallic Josephson Junctions}
\label{jojo_results}
\begin{figure}
\includegraphics
[width=3.4in]
{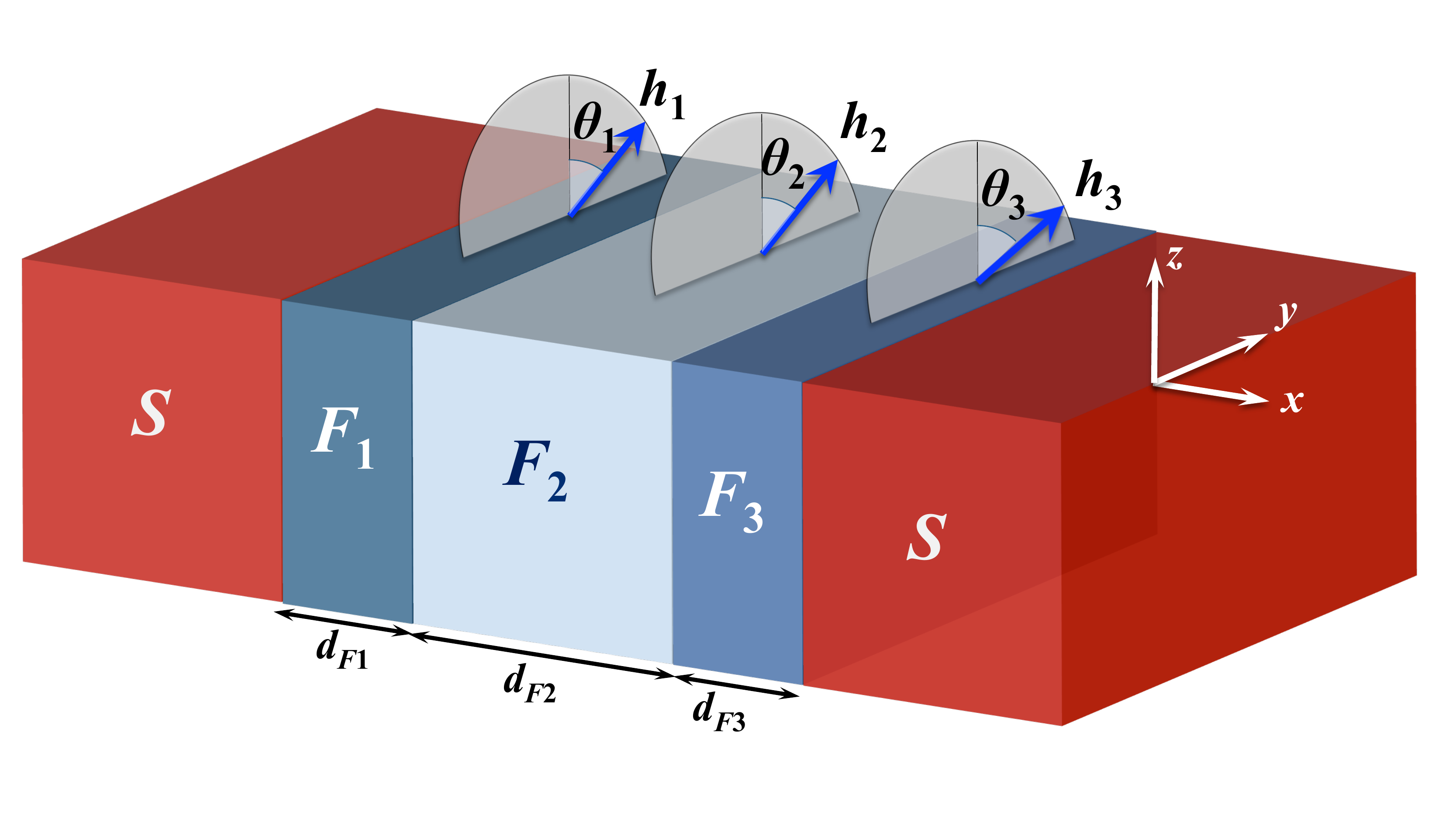}	
\caption{
(Color online) Schematic of the $S F_1F_2 F_3 S$ Josephson junction. The layers are  %kh
translationally invariant and extend to infinity in 
 the $yz$ plane.
The central $F_2$ layer is half-metallic ($h_2=E_F$), while the surrounding $F_1$ and %kh
$F_3$ layers are ferromagnets with weaker exchange fields  %kh
$h_1=h_3=0.1 E_F$. The angles $\theta_1$, $\theta_2$, and $\theta_3$ describe the 
angles that the magnetic exchange field vector makes with the $z$ axis in the corresponding 
$F_1$, %kh
$F_2$ and $F_3$ layers with thicknesses $d_{F1}$, $d_{F2}$, and $d_{F3}$, respectively. %kh
}
\label{diagram}
\end{figure}
In this subsection we present  our results 
for half-metallic 
$S F_1F_2 F_3 S$ Josephson  junctions.
A diagram of the setup is shown in Fig.~\ref{diagram}.
A trilayer magnetic configuration is considered to allow for 
the generation of singlet and triplet correlations by using relatively weak and thin
magnets nearest the $S$ layers.
For the half-metal thicknesses considered here,
using a simpler bilayer structure consisting of a thick half-metal and ferromagnet
would result in the destruction of phase coherence between the $S$ banks.
Thus two relatively weak ferromagnets are needed to be in contact with the superconductors 
to effectively generate  triplet correlations and 
establish both charge and spin currents within the junction.
The thicknesses of the $S$ layers are $800/k_F$,
while
$F_1$, $F_2$, and $F_3$ can vary,
depending on the quantity being studied.
As before,
the superconducting coherence length is fixed to be $100/k_F$.
For most cases, the interfaces are generally assumed to be transparent, 
although cases with interface scattering will be considered as well.
Unless otherwise noted, the central $F$ layer is
half-metallic,
with exchange field corresponding to $h_2=E_F$. 
Similar to what was shown for tunnel junctions,
the spin-valve effect is maximized when the exchange fields of
$F_1$ and $F_3$ are weaker: We consider here 
$h_1=h_3=0.1E_F$. 
For these  Josephson structures, 
the focus of the investigation is on the influence that the
 macroscopic phase difference $\Delta\varphi$, and the relative 
 magnetization orientations
 have on the spin currents, charge currents, and
associated triplet correlations.
To be consistent with the previous results on tunnel junctions, the magnetization is normalized
by $-\mu_B n_e$, where $n_e$ is the electron density %: $n_e = k_F^3/(3\pi^2)$,
and the charge currents are normalized 
by $J_0$, where $J_0=e n_e v_F$, and $v_F = k_F/m$ is the Fermi velocity.
All three components of the spin current $\bm S$ are normalized
similarly \cite{hvw15}.

\begin{figure}
\includegraphics
[width=3.2in]
{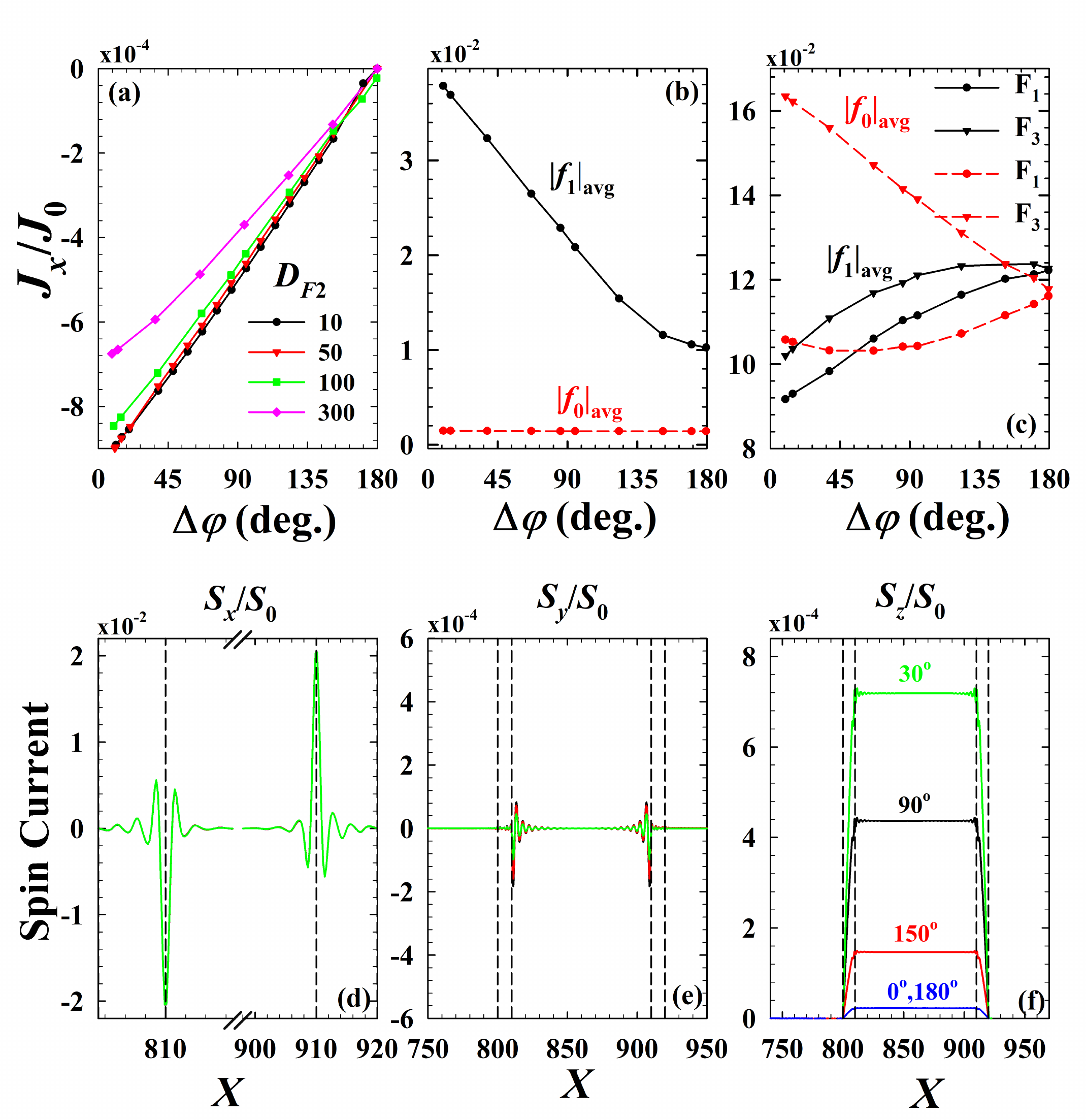}
\caption{
(Color online)
First row (a)-(c):
The normalized 
current density $J_x$ v.s. 
the phase difference $\Delta\varphi$ for
(a) several normalized half-metal thicknesses $D_{F2}=k_F d_{F2}$.
(b) The equal-spin $f_0$ and opposite-spin $f_1$ triplet correlations spatially averaged over 
the half metal region $F_2$, and  (c) over the  ferromagnetic  $F_1$ and $F_3$ regions.
Second row (d)-(f): The components of the normalized spin current $\bm S$ 
as a function of dimensionless position $X=k_F x$.
In (a)-(f), the  exchange fields in
$F_1$ and $F_3$ are aligned along the $y$ direction, and 
along $z$ in the half-metal $F_2$ (see Fig.~\ref{diagram}).
The ferromagnets  $F_1$ and $F_2$ have equal thicknesses of
$10/k_F$.
In panels (b)-(f) the $F_2$ thickness  is fixed at 
$100/k_F$.
}
\label{cpr_10_x_10}
\end{figure}
We begin with the self-consistent current phase relation for
the $S F_1F_2F_3 S$ structure shown in Fig.~\ref{diagram}.
In Fig.~\ref{cpr_10_x_10}(a),
 the normalized charge current flowing in the $x$ direction, $J_x$, is shown  as a function of the
 macroscopic phase difference
$\Delta\varphi$.
The central half-metallic $F_2$ layer is 
sandwiched between two weaker ferromagnets with normalized 
exchange field strengths $h/E_F=0.1$, 
and thicknesses  $10/k_F$.
Each of the ferromagnets $F_1$ and
$F_3$ have their magnetizations oriented
in the same direction (along $y$) but orthogonal to $F_2$ (along $z$).
To isolate the triplet spin current flowing through  the half-metal, 
differing dimensionless thicknesses $D_{F2}=k_Fd_{F2}$ are considered, as 
shown in the legend. 
As seen, the supercurrent essentially 
obeys a linear trend with phase difference that is weakly  dependent on
the thickness of the half-metal. As this thickness increases, 
 the current begins to deviate from the
linear behavior, as seen developing for the $D_{F2}=300$ case.
The fact that increasing the thickness $D_{F2}$ has a weak effect 
on the supercurrent reflects the spin-polarized nature of the triplet pairs involved in transport
through the half-metal.
We limit the range of the current phase relation for clarity, 
however extending the range of $\Delta \varphi$ 
would result in a sawtooth-like profile with vanishing 
current at $\Delta\varphi=n\pi$, where is $n$ is an integer. 
Physically, the slow decay of the equal-spin triplet correlations in the half metal
equates to propagation lengths of the quasiparticles that can well exceed $\xi_F$. 
To demonstrate  this, in (b) the magnitudes of the
 opposite spin
correlations $f_0$ and equal spin correlations $f_1$
averaged over the half-metallic region $F_2$
are shown.
To satisfy the Pauli principle, these
spatially symmetric triplet pairing correlations must be
odd in  time, and hence vanish when the relative time $t$ is zero.
For the results involving  triplet pairs in this section, we
take the corresponding dimensionless time to be  $\omega_D t=4$. %kh check
Due to the presence of only one spin band in  $F_2$, the $f_0$ correlations
have a very weak extent within the half-metal and remain relatively constant for all $\Delta\varphi$.
On the other hand, the $f_1$ component has a relatively large presence in $F_2$,
increasing as the magnitude of the current increases. In the absence of current,
the triplet amplitudes populate the half-metal,
consistent with what is found in half-metallic spin valves~\cite{half}.
As mentioned earlier, the presence of the thin ferromagnet layers is important
for the generation of the opposite-spin triplet pairs, and consequently the conversion to the
equal-spin channel. This effect is clearly seen in (c), where now the 
magnitude of the triplet correlations are presented averaged over the $F_1$ and $F_3$ layers.
As the macroscopic phase difference
 changes, it is evident that a nontrivial intermixture of $f_0$ and $f_1$ occurs in those layers.

In the bottom row of panels ((d)-(f)), the three components of the normalized spin currents are shown
as a function of the dimensionless position $X=k_F x$.
All components of the spin current flow in the $x$ direction. 
The dashed vertical lines serve to
identify the narrow ferromagnetic regions containing $F_1$ and $F_3$.
If the $F$ layers possessed  uniform magnetization, there would be no
net spin current. The introduction of an inhomogeneous magnetization
 however results in a net spin current imbalance
that is finite even in the absence of a Josephson current.
In (d), we present 
the normalized $x$ component of the spin-current,  $S_x$, which 
is responsible for the torque that tends to align the 
magnetizations in the ferromagnetic layers.
This exchange field mediated effect is present 
in the absence of
Josephson current and is seen to be almost independent  of
the phase difference that drives the Josephson current.
As seen, this quantity is maximized at the interfaces, before undergoing damped oscillations.
For completeness, we have included in (e) the $y$ component of the spin current,
which for our magnetic configuration is clearly negligible.
In panel (f), we examine the normalized  $z$-component of the spin current $S_z$.
This component, which is oriented parallel to the interfaces tends to build up on the weakly ferromagnetic layers
and then propagate uniformly in the half-metal. The magnitude of $S_z$ is seen to correlate with the magnitude of
the charge current in (a), where 
the smaller phase differences result in 
large charge and spin currents that decline as $\Delta\varphi$ increases. 
These results indicate that the half-metal polarizes the spin current along its magnetization 
direction, and that the 
Josephson current is due to the propagation
of equal-spin triplet pairs.

\begin{figure}
\includegraphics
[width=3.4in]
{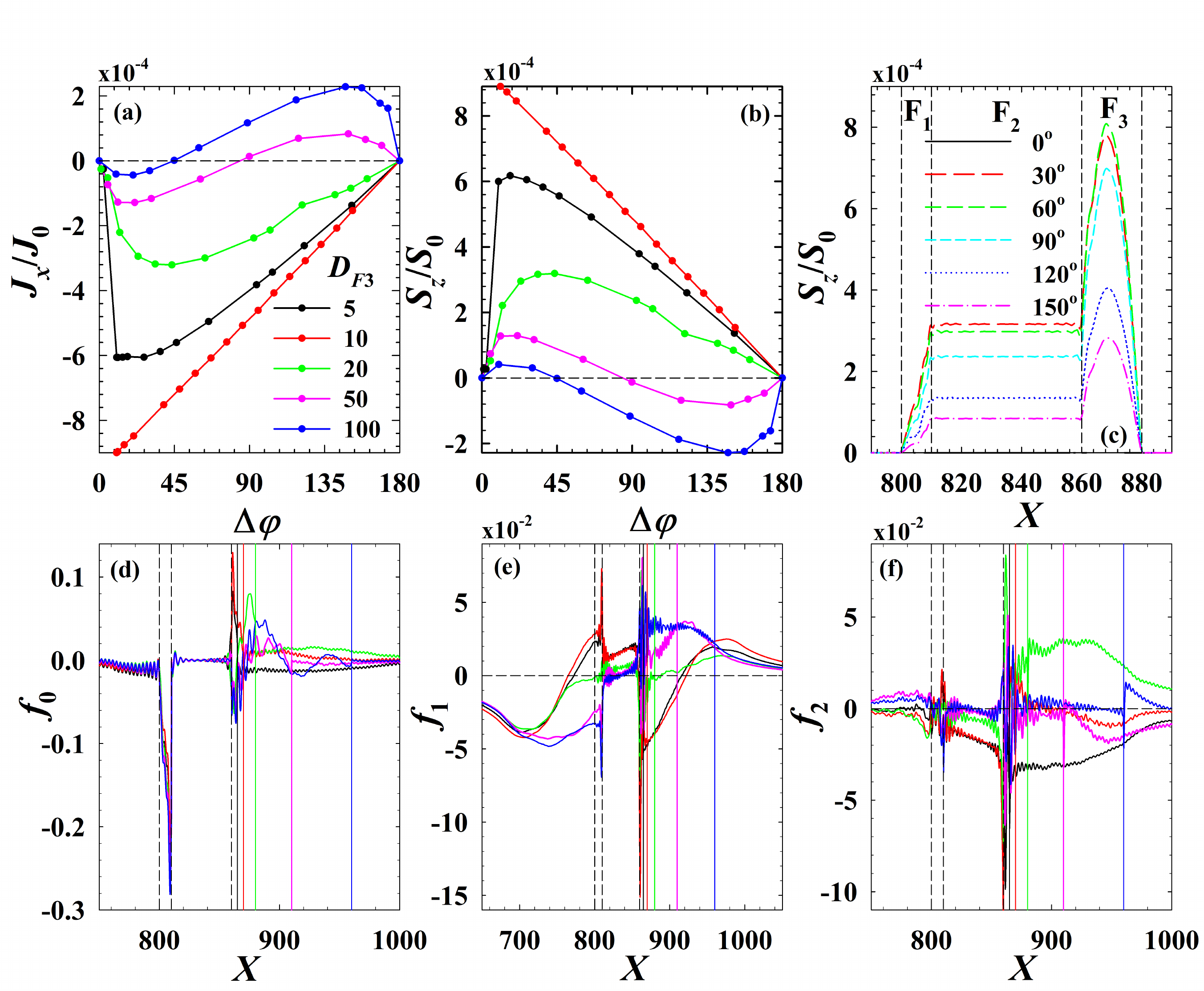}
\caption{
(Color online)
Top row: (a)
The normalized charge current density 
$J_x$ v.s. the phase difference $\Delta\varphi$,
(b) the $z$ component 
of the normalized spin current density $S_{z}$ within the half-metal region v.s. $\Delta\varphi$
for several $D_{F3}$,
and (c) the normalized  $S_{z}$ as a function of dimensionless position $X\equiv k_F x$
for $D_{F3}=20$. The legend in (c) labels the 
different phase differences $\Delta\varphi$ (in degrees) between the $S$ banks.
%The quantity $S_{z,{\rm avg}}$
%is obtained by averaging $S_z$ over the half-metal region.
The legend in (a) depicts the ferromagnet thicknesses $D_{F3}$ used in (a), (b), and (d)-(f).
Bottom row:
The spatial behavior of the real part of the triplet correlations for various 
 thicknesses (see legend in (a)) and for a phase difference of $\Delta\varphi=90^\circ$.
The dashed vertical lines identify the $F_1$ and $F_2$ regions located within
$800\leq X\leq 810$ and $810< X \leq 860$ respectively, while the solid vertical lines mark 
 the various $F_3/S$ interfaces.
The  exchange field in
$F_1$ and $F_3$ is aligned along the $y$ direction, while it points along $z$
in the half-metal (see Fig.~\ref{diagram}).
The thicknesses $d_{F1}$, and $d_{F2}$ are
maintained at the constant dimensionless values of 
$D_{F1}=10$ and $D_{F2}=50$, respectively.
}
\label{cpr_10_50_x}
\end{figure}
Next, in Fig.~\ref{cpr_10_50_x}(a)
the half metal $F_2$ and ferromagnet $F_1$
have  fixed thicknesses 
corresponding to $D_{F2}=50$ and
$D_{F1}=10$, respectively.
The
ferromagnet $F_3$ is
allowed to vary, as shown in the legend.
Asymmetric structures with unequal thicknesses of the ferromagnetic layers
has been shown to enhance spin mixing effects that results in 
the generation of long-ranged spin-polarized triplet pairs~\cite{zep}.
The  linear behavior of the charge current previously shown
in Fig.~\ref{cpr_10_x_10} where 
the two magnets $F_1$ and $F_3$
are of equal thickness is seen to
transition to a sinusoidal-like structure as the difference in 
the thicknesses between $F_1$ and $F_3$ increases. 
Thus, for highly asymmetric structures, the current phase relation
reveals a sign change in the charge current for phase differences between $0^{\circ}$
and $180^{\circ}$.
The  ferromagnet  $F_3$ with relatively weak exchange field 
compared with $F_2$
and somewhat larger thicknesses ($D_{F3}=10$)
creates ideal conditions 
 for the  creation and propagation  of
 opposite-spin triplet pairs.
The center of mass momentum of a given pair shifts 
in the presence of spin splitting from the exchange field,
resulting in the observed damped oscillations for a given
$\Delta\varphi$.

If we now calculate the $z$ component of the spin current flowing through the half-metal
portion of the junction, we find that aside from a sign difference, it is nearly identical to the Josephson current
as seen in Fig.~\ref{cpr_10_50_x}(b). %kh
This reaffirms that the current flowing through the half-metal is comprised 
of Cooper pairs that are polarized in the $z$ direction by
 the half-metal.
 In general, the spin current is a non-conserved quantity, in contrast to the charge current.
Thus, although  $S_z$ is uniform throughout the half-metal, it spatially varies in the other junction regions.
 This is demonstrated in (c)
 for several phase differences $\Delta \varphi$ (see legend), 
 where $D_{F1}=10$, $D_{F2}=50$,
 and $D_{F3}=20$.
 The spin current does not flow in the outer superconductor banks, and thus
 $S_z$ increases from  zero 
  at the $S/ F_1$ interface ($X=800$)
 before 
 reaching its uniform value in the half metal, and then
  peaks within $F_3$ 
  before declining to zero again in the superconductor.

 To reveal the relative population of
    triplet pairs  throughout the junction, 
    we consider in (d)-(f)
    the triplet correlations $f_0$, $f_1$, and $f_2$,
    as functions of normalized position $X$. The
   phase difference is set according to   $\Delta \varphi=90^\circ$.
    We still have $D_{F1}=10$, and $D_{F2}=50$, but several
    $D_{F3}$ are shown with values given in the legend found in panel (a), 
  thus creating 
  a broad range of current profiles.
The opposite-spin triplet correlations  shown in (d) reveal that
$f_0$ spikes in the $F_1$ region, weakly dependent on $D_{F3}$.
Within $F_2$ however, 
the single spin band present in the half-metal severely 
diminishes $f_0$. 
When 
 $F_3$ 
 has thin layers,
 the greater confinement 
 enhances the $f_0$ amplitudes.
 Increasing $D_{F3}$ eventually 
 provides sufficient space for  the exchange field 
to induce damped oscillations
of the opposite-spin pairs.  
Thus, although it is energetically unfavorable for the
$f_0$ correlations to 
reside in the half metal, 
they do become enhanced in %kh
the surrounding ferromagnets when they are thin %kh
($D_{F3}=5,10$).
Under these conditions, the spin polarized triplet pairs
$f_1$ and $f_2$ propagate within the half metallic region,
as seen in (e) and (f).
It is also evident that often 
 the equal-spin triplets do not decay within the $S$ regions, but rather
 extend deep into the superconductor banks.

\begin{figure}[t!]
\includegraphics
[width=3.2in]
{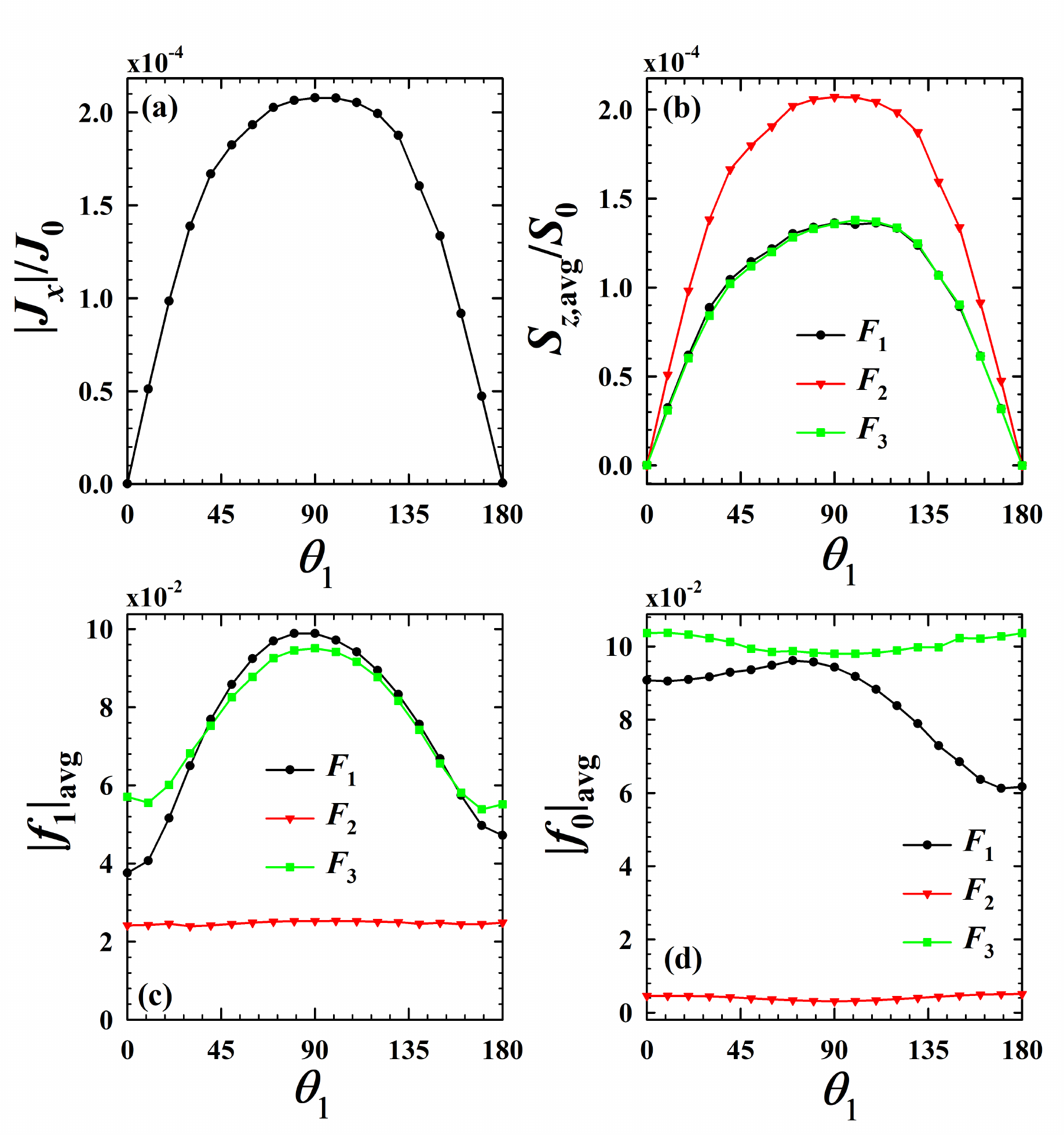}
\caption{(Color online) Top row: (a) Normalized charge currents and (b) the $z$ component of normalized spin currents in each 
junction region as a function of the magnetization alignment angle $\theta_1$. %ct
Bottom row: Spatially averaged equal-spin (c) and opposite-spin (d) triplet correlations as a function of $\theta_1$.%ct
The thicknesses of ${F_1}$, ${F_2}$, and ${F_3}$ are set to be
$10/k_F$, $50/k_F$, and $10/k_F$, respectively. An interface scattering strength of $H_{1,4}=0.8$ is present at the interfaces (see main text),
and a phase difference of $\Delta\varphi=90^\circ$ is assumed.
}
\label{current_phi1}
\end{figure}
Having seen the influence that the layer thicknesses in half-metallic Josephson junctions have on the charge and spin currents, 
we now turn to the
the effects of  magnetization rotations.
Rotating  the magnetization in one of the junction layers
can be achieved experimentally via external magnetic
fields, or spin-torque switching.
 In Fig.~\ref{current_phi1}(a) we display 
the magnitude of the normalized charge current as a function of the magnetization angle $\theta_1$. %ct
The half metal thickness is set at $D_{F2}=50$, and the surrounding ferromagnets 
have equal thicknesses of $D_{F1}=D_{F3}=10$.
The effects of scattering at the $S/F_1$ and $F_3/S$ interfaces are
accounted for by setting the dimensionless parameter 
$H_{B1}\equiv H_1/v_F=0.8$ and $H_{B4}\equiv H_4/v_F=0.8$, respectively.
Here $H_1$ and $H_4$ are the delta-function scattering strengths  at those two %kh
interfaces~\cite{hvw15}.
The inclusion of interfacial scattering in Josephson junctions tends to suppress the linear sawtooth profile in
the current phase relation~\cite{hvw15}.
The Josephson current is
established with a phase difference $\Delta\varphi=90^\circ$ between the superconducting banks.
The half metal layer has its ferromagnetic exchange field directed along $z$ and  for $F_3$,
it is directed  along $y$ (see Fig.~\ref{diagram}).
Thus, when $\theta_1=0^\circ$  or $\theta_1=180^\circ$, both $F_1$ and the adjacent half metallic layer have magnetizations that are parallel %ct
or antiparallel, respectively. 
At these points, $J_x$ vanishes while  
the supercurrent flow is largest when $\theta_1=90^\circ$, corresponding to  %kh2
when the junction layers have magnetizations that are orthogonal to one another, %kh
and hence possess a high degree of magnetic inhomogeneity.
The half-metal tends to align the spin of any entering quasiparticles 
along the $z$ direction,
and this component of the normalized spin current 
displays nearly identical behavior to $J_x$ as seen in (b).
The averaged spin current  
is  distributed equally throughout the two outer ferromagnets,
but weaker overall  since
it must vanish at the boundaries with the superconductors.
The behavior of the magnitudes of the triplet correlations v.s. $\theta_1$ %ct
 is presented in panels (c) and (d).
When $\theta_1=0^\circ$  or $\theta_1=180^\circ$, the generation of  equal-spin %ct
triplets are suppressed in the ferromagnets $F_1$ and $F_3$
due to the lowering of the overall magnetic inhomogeneity. For these situations,
 the magnetizations in the $F_1$ and $F_2$ layers
are collinear, however, $f_1$ does not vanish due to the orthogonal magnetization  in $F_3$.
On the contrary, when $\theta_1=90^\circ$, the magnetization in each %ct
ferromagnet is orthogonal to the adjacent one, resulting in favorable conditions for the creation of
the equal spin triplets. 
In (d) the importance of having relatively weak and thin outer ferromagnets for the triplet conversion process is exhibited by the 
population of the $f_0$ triplet components in those regions.

\begin{figure}
\includegraphics
[width=3.2in]
{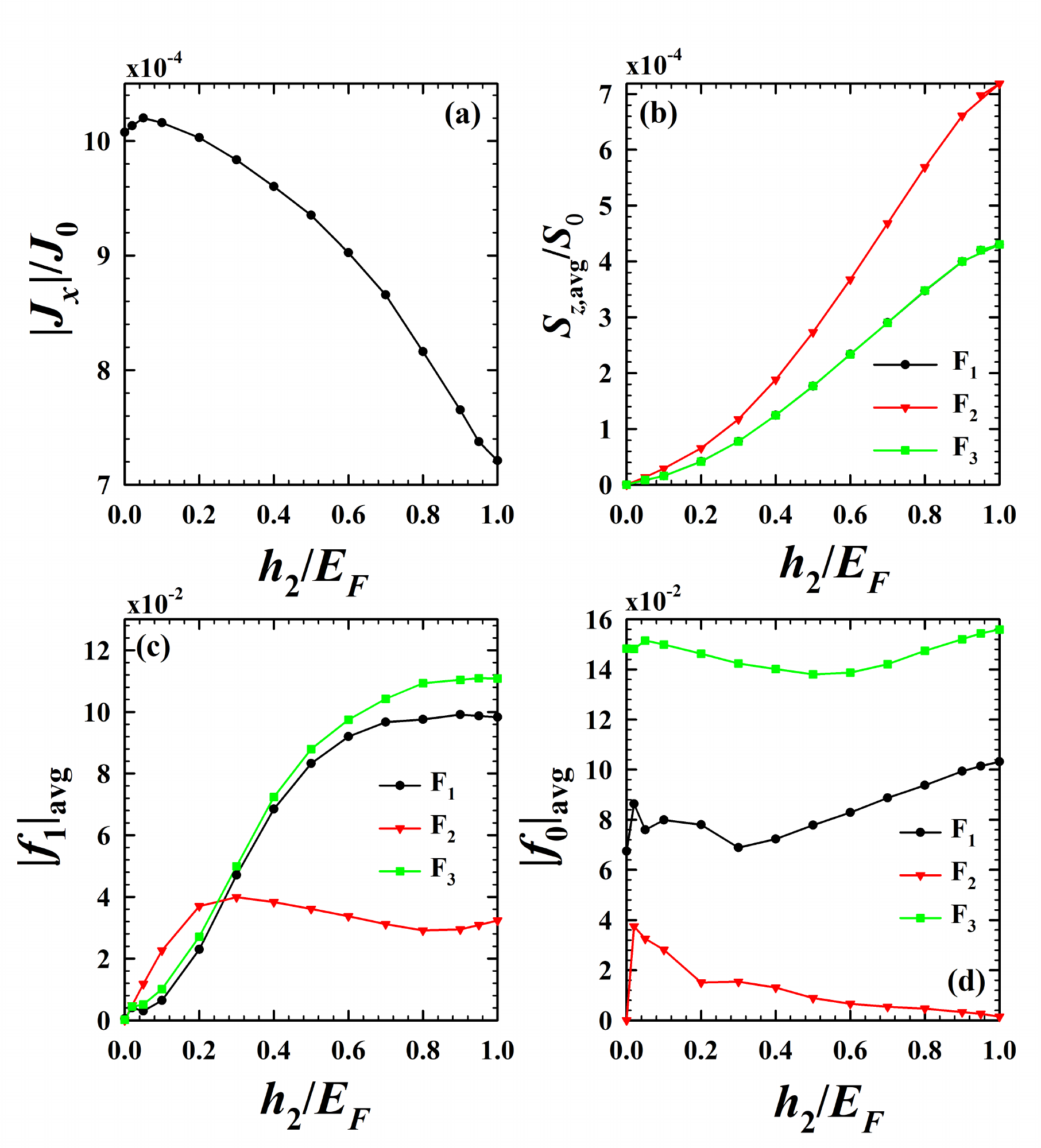}
\caption{
(Color online) Top row: Normalized charge current (a) and average spin currents (b) v.s. the dimensionless
magnetization strength $h_2/E_F$.
Bottom row: Spatially averaged equal-spin (c) and opposite-spin (d) triplet correlations as a function of $h_2/E_F$.
The thicknesses of $F_1$, $F_2$, and $F_3$ are set to be
$10/k_F$, $100/k_F$, and $10/k_F$, respectively. 
The legend in (b) identifies each region of the junction in which the quantities in (b)-(d) are averaged over.
Here, transparent interfaces are considered and $\Delta\varphi=30^\circ$.
}
\label{currenth}
\end{figure}
It was observed that the presence of the half-metal in the junction serves to filter out 
the opposite-spin triplet pairs, creating a platform in which to study
spin polarized triplet correlations.
It is of interest to clarify
 the role that the exchange field strength in the half metal region has on the charge and spin transport.
The top row of Fig.~\ref{currenth} therefore 
shows the magnitude of the charge current and the averaged spin current, both normalized, 
as a function of the exchange field strength in the half metal, $h_2$. The phase difference is
set to $\Delta\varphi=30^\circ$. For clarity, the two ferromagnets have equal thicknesses,
$D_{F1}=D_{F3}=10$, and there is no interface scattering present.
The larger half metal has a thickness of $D_{F2}=100$, and the exchange field varies from $h_2=0$ to $h_2=E_F$,
which coincides with a nonmagnetic normal metal and a half-metallic phase, respectively. 
The junction's magnetization profile %kh
is in an optimal inhomogenous state, with alignment angles are as follows: $\theta_1=90^\circ$, $\theta_2=0^\circ$, %ct
and $\theta_3=90^\circ$, %ct
corresponding to magnetization alignments along  $y$, $z$, and $y$, respectively. 
Examining panel (a), it is evident that the magnitude of the 
charge current $J_x$ is maximal
when the $F_2$ layer is weakly ferromagnetic, and is minimal  when $F_2$ is
half-metallic.  
The spatially averaged spin current 
on the other hand is anticorrelated with $J_x$,
as it monotonically increases with larger exchange fields. 
Indeed, $S_z$ vanishes when the central $F_2$ 
layer is a nonmagnetic normal metal, and peaks when it is half-metallic.
When the central layer is nonmagnetic $S_z$ vanishes since
the only active magnetic layers in this case are $F_1$ and
$F_3$ which have parallel magnetization directions.
Examining the bottom row, 
the triplet correlations are also shown 
averaged over each of the three junction layers. 
In (c) the magnitude of the $f_1$ correlations are shown v.s. $h_2/E_F$.
When $h_2=0$, $F_1$ and $F_3$ are the only ferromagnetic layers in the junction,
and their
magnetizations are 
oriented along $y$. 
Since they are collinear,
spin-polarized triplet pairs cannot be generated, and hence $f_1=0$.
Increasing $h_2$ and hence the degree of polarization in the $F_2$ layer
continuously increases the amount of spin polarized triplet pairs in the 
ferromagnets $F_1$ and $F_3$, with $f_1$ largest when
$F_2$ is  half-metallic. %kh
The $f_1$ correlations in $F_2$  also
become enhanced as its exchange field get larger, until $h_2/E_F\approx0.3$.
Further increases in $h_2$  result in  a slight decline before ultimately 
increasing again as $F_2$ approaches the half-metallic limit.
This demonstrates the importance of using a highly spin-polarized material in the central
junction region to optimize triplet pair generation in each layer.
The opposite-spin pairs $f_0$ are also maximized 
in the triplet conversion layers $F_1$ and $F_3$ %kh
when $h_2=E_F$ as seen in (d).
Unlike what is found for $f_1$, the $f_0$ correlations are not constrained to vanish when $h_2=0$
since they can exist when the ferromagnets have collinear magnetizations. Thus the thin ferromagnetic regions
have a substantial portion of $f_0$ pairs when $h_2=0$.
Within the thicker $F_2$ layer however, $f_0$ is significantly reduced overall, 
becoming negligibly small  in the nonmagnetic metal ($h_2=0$) and half-metallic ($h_2=E_F$)
limits. It substantiates the idea that using the half-metal for $F_2$ enables us
to focus on the interplay between the spin current and the equal spin pairs $f_1$  
in the $F_2$ region.

\begin{figure}
\includegraphics
[width=3.4in]
{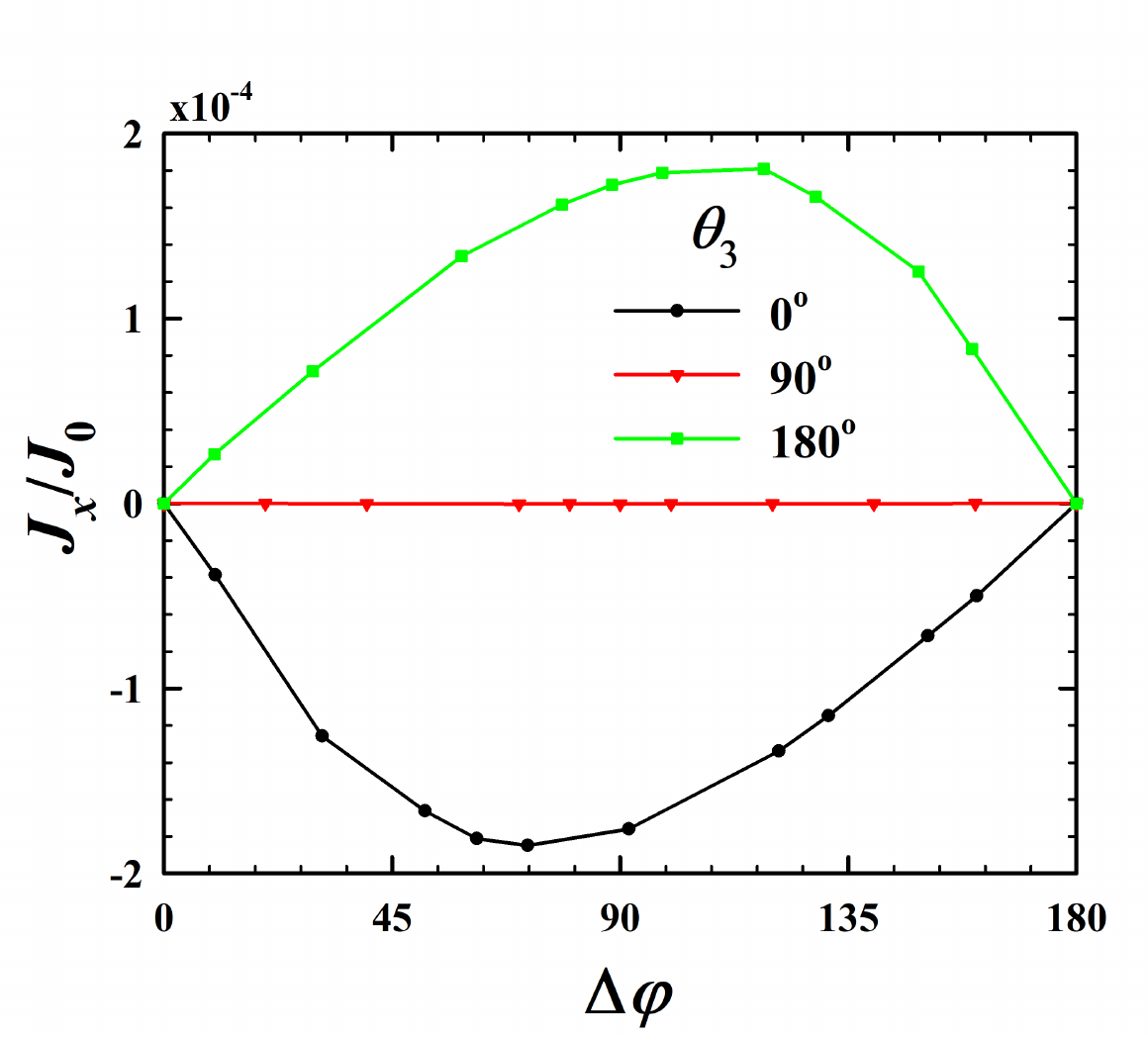}
\caption{
(Color online) Charge current %ct (a) and triplet correlations (b) and (c)
 as a function of
phase difference $\Delta \varphi$.
The thicknesses of $F_1$, $F_2$, and $F_3$ are set to be
$10/k_F$, $100/k_F$, and $10/k_F$, respectively. The interface scattering strengths 
are set to $H_{B1}=H_{B4}=0.8$. The magnetization in $F_1$ and $F_2$ is along the
$z$ and $y$ directions respectively. In the ferromagnet $F_3$, the magnetization orientation angles
%correspond to $\theta_3=90^\circ$ and 
$\theta_3$ varies as shown ($\theta_3=0$ is along $z$, $\theta_3=90^\circ$ is along $y$, and $\theta_3=180^\circ$ is along -$z$). %ct
}
\label{cpr2}
\end{figure}
We now take the structure previously studied above
in Fig.~\ref{currenth} and incorporate interface scattering, and 
rotate the 
 magnetizations so that they are interchanged 
for the first two layers. Thus,  $F_1$ and $F_2$ 
have their magnetizations aligned along 
the $z$ and $y$ axes respectively.  
The normalized interface scattering strength is set at $H_{1}=H_{4}=0.8$.
With these parameters, 
Fig.~\ref{cpr2} examines the normalized Josephson supercurrent 
as a function of the phase difference $\Delta\varphi$.
Three magnetization orientations 
 for $F_3$ are investigated for each of the three panels:
$\theta_3=0,90^\circ$, and $180^\circ$ (corresponding to the $z$, $y$, and  $-z$ directions, respectively).
The  supercurrent  reveals that,
 depending on whether 
the magnetization in $F_3$ is collinear or orthogonal to the adjacent half-metal,
the direction of the charge current can be reversed  or turned off completely. 
When $\theta_3=0$, the magnetization in each layer is orthogonal to one another, and  the current phase relation  %kh2
reveals that when starting from zero phase difference, the magnitude  of the current 
increases until $\Delta\varphi\approx 70^\circ$, before declining back to zero again at $\Delta\varphi=180^\circ$.
Due to quasiparticle scattering that takes place at the interfaces, 
the coherent transport of Cooper pairs through the junction
is significantly altered compared to when the interfaces were transparent, 
resulting in the observed overall reduction in current and
deviation from the previous linear behavior found in Fig.~\ref{cpr_10_x_10}.
Previously, when  studying how magnetization rotation affected the charge current
in Fig.~\ref{current_phi1}(a),
 we found that 
 when two adjacent layers in the junction have collinear  magnetizations,
the charge current vanished. 
This is consistent with 
Fig.~\ref{cpr2}, where the current vanishes for all phase differences  at $\theta_3=90^\circ$. %kh2
Rotating the magnetization further to $\theta_3=180^\circ$,  %kh2
the magnetizations in 
both ferromagnets are orthogonal to the half-metal, as in the $\theta_3=0$ case,
but antiparallel to each other. This causes a reversal of the charge current as shown.

\begin{figure*}
\includegraphics
[width=5.2in]
{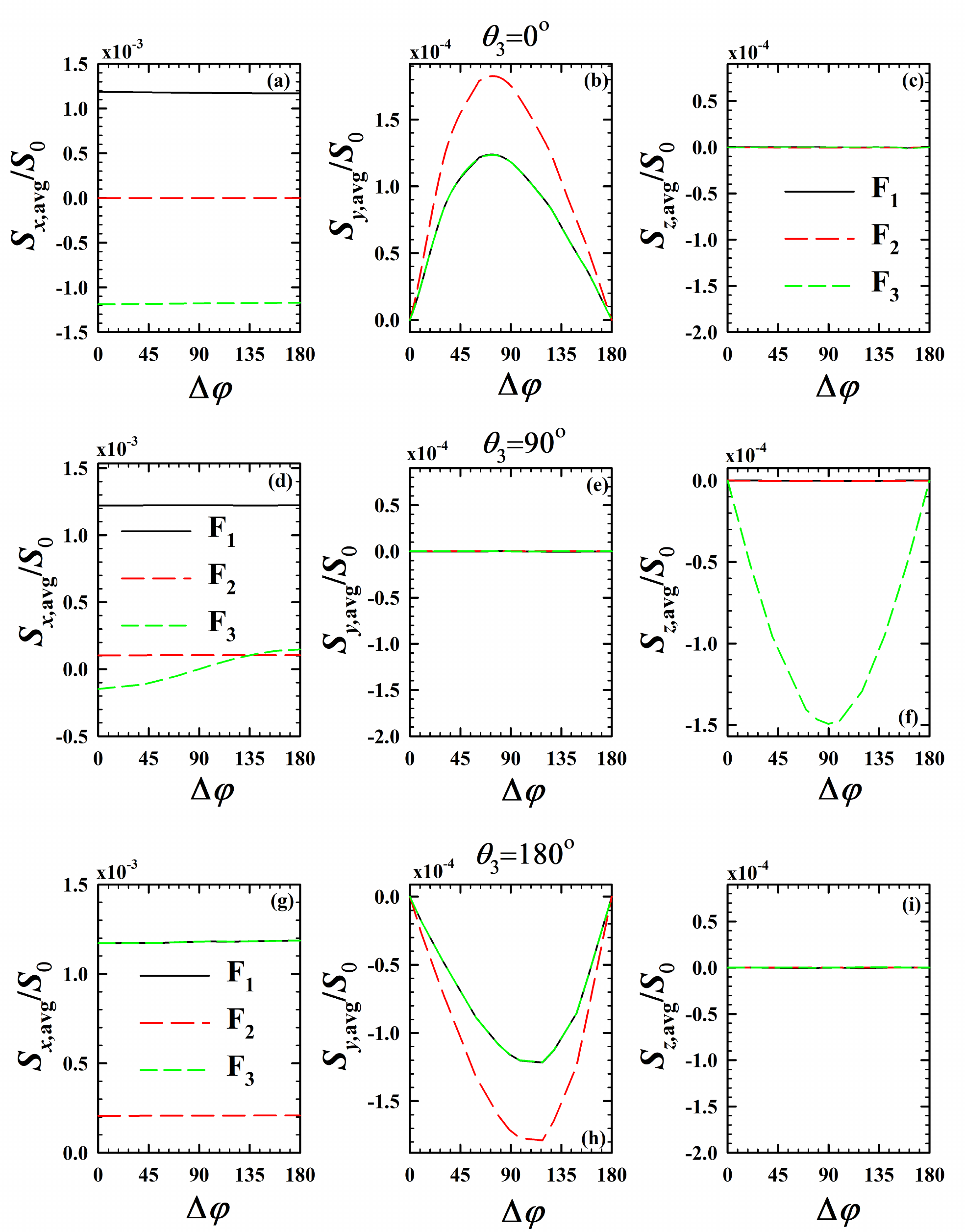}
\caption{
(Color online)
Components of the average spin current in each region v.s. the phase difference $\Delta\varphi$.
Three magnetization orientations of the outer ferromagnet $F_3$ %kh
are shown: $\theta_3=0$ (top row), $\theta_3=90^\circ$ (middle row), %ct
and $\theta_3=180^\circ$ (bottom row). %ct
The thicknesses of $F_1$, $F_2$, and $F_3$ are set to be
$10/k_F$, $100/k_F$, and $10/k_F$, respectively. The dimensionless interface scattering strengths
correspond  to $H_{B1}=H_{B4}=0.8$.
}
\label{spin_10_100_10}
\end{figure*}
As shown earlier, the charge current that flows due to
the macroscopic phase differences between the $S$ electrodes
can become spin-polarized when 
entering one of the ferromagnetic or half-metal layers. 
This  spin current can then interact with
the other ferromagnets and become modified by the corresponding magnetizations.
Having established in Fig.~\ref{cpr2} how the charge current can be
manipulated  for a half-metallic
Josephson junction, it is important to next identify %ct
how the spin currents behave in each layer, as
control of these spin currents is vital for spintronic applications.
Thus, in Fig.~\ref{spin_10_100_10} we
investigate  the phase dependence  for the spatially averaged spin current 
 $\bm S$.
 We implement  the same experimentally accessible parameters used in
 Fig.~\ref{cpr2}. Each row of three panels corresponds to one of the
  three  magnetization orientations $\theta_3$ (as labeled). %ct
As discussed  earlier, 
the central half-metallic layer maintains a constant 
spin current,
 that can  couple the surrounding ferromagnets $F_1$
and $F_3$. %ct
 This effect is evident for  $S_x$ when $\theta_3=0^\circ$ (top row), %kh2
corresponding to the $z$,$y$,$z$ magnetic configuration 
for the respective $F_1$, $F_2$, and $F_3$ layers.
This spin current component, normal to the interfaces,
is essentially 
the static contribution to the spin current, which 
participates in  spin-transfer torque effects 
near the ferromagnet/half-metal interfaces where misaligned
exchange fields are present. 
Thus,  $S_x$  varies in space, 
resulting in a local STT [recall $\partial S_x /\partial x = \tau_x$] 
that tends to rotate the corresponding magnetizations in opposite directions.
Within the half-metal, the spin current oscillates as it damps out deep within $F_2$,
resulting in an average $S_x$ of zero, as  exhibited in (a).
The averaged spin currents clearly do not depend on the phase difference,
as expected for a static effect. 
The strong influence of the half-metal is exhibited by 
$S_y$, the spin current  
component that lies in the same direction
as the exchange field in the half-metal. 
The half-metal is seen to polarize not only the spin current
within it, but also within  the surrounding weak magnets whose intrinsic exchange fields
are in  the orthogonal $z$ direction. 
Note that the $y$-component of spin currents in each of the $F$ regions 
have
similar  overall behavior as a function of $\Delta\varphi$,
with the average $S_y$ being equal in $F_1$ and $F_3$,
and largest in $F_2$. %ct
Comparing this to Fig.~\ref{cpr2}, it is clear that apart from a sign difference, the normalized
 spin current $S_y$ in the half-metal  
and the supercurrent $J_x$ are nearly identical. This implies that
the spin-polarized current $S_y$ in the half-metal
correlates with the charge current that is flowing there. Therefore,  the 
charge transport is governed 
by  spin-polarized Cooper pairs 
corresponding to the  equal-spin correlations. 
Turning now to the middle row,
where $\theta_3=90^\circ$, there is no spin current along $y$ for all of the $F$ layers. %ct %kh2
Within the $F_3$ layer, the normalized $S_x$ is
shown to vanish at $\Delta\varphi=90^\circ$, while $S_z$ is maximal for that phase difference.
Considering the phase differences that yield no supercurrent, $\Delta\varphi=0^\circ$ and $180^\circ$,
The spin currents $S_x$ and $S_z$ are seen to be anti-correlated, with
$S_z$ now  vanishing, and the magnitude of $S_x$ having now become largest in $F_3$.
Finally, the bottom row depicts the spin currents for $\theta_3=180^\circ$. %ct
As was found for the previous $\theta_3=0^\circ$ case, %ct
we see a direct correlation between the charge supercurrent [Fig.~\ref{cpr2}] and 
the $y$ component of the spin current for this magnetic configuration.
The main differences being that the directions of the charge and spin currents are reversed,
due to $\theta_3$ having a reversed collinear orientation, %ct
and non-vanishing $S_x$ in $F_2$. %ct

\section{Conclusions}
\label{conclusion}
In this paper, we have studied in detail the interplay between the
triplet pairs and transport properties of half-metallic 
superconducting spin valves including tunnel junctions
and Josephson junctions. In tunnel junctions 
with the presence of an applied bias voltage,
we have discussed a useful theoretical approach
combining the self-consistent solutions to the 
Bogoliubov-de Gennes equations and
the transfer matrix method based on 
the Blonder-Tinkham-Klapwijk formalism. 
By utilizing this approach, we are able to 
determine the bias dependence
of spin transport quantities and the
induced triplet pair amplitudes.
We first investigate the bias-induced
magnetizations, spin currents, and 
the spin-transfer torques 
as functions of position for
various misorientation angles %kh
between the half metal and 
adjacent weak ferromagnet. 
We find that their behaviors
can be largely explained by
the precessional effect: When the injected charge 
current spin-polarized by the half-metal enters
the weak ferromagnet, its polarization
state can be rotated by 
the local exchange interaction.
The bias dependence of these
spin transport quantities
are also studied. We find that
their magnitudes increase linearly 
when external bias voltages 
are larger than the saturated superconducting
pair amplitudes.
We then show that
the spin transport quantities
are determined by two important
parameters: the exchange interaction
and thickness of the weak ferromagnet.
Both $m=0$ and $m=\pm 1$ triplet correlations
of the tunnel junctions are also 
presented. We find that they are anti-correlated
when the misorientation angle between exchange interactions %kh
in the ferromagnetic layers is varying.
Furthermore, the long-range nature 
of $m=\pm 1$ triplet correlations
in the half-metallic region
is established and proven 
to be important in the half-metallic tunnel junctions.
It is shown that
choosing the exchange interaction to be about $0.1E_F$
can optimize the spin-valve effect.
We then switch to half-metallic Josephson junctions which
consist of a half metal sandwiched by two weak ferromagnets
in the non-superconducting region.
First, we consider a symmetric situation where the thicknesses
and the exchange fields are the same for two weak magnets.
To generate all components of triplet pairs, the
exchange field in the half metal is perpendicular to 
that of weak magnets.
We study the current phase relations and find that 
the current is only weakly dependent on the thickness
of the half metal indicating that
the supercurrent is carried by the equal-spin triplet pairs.
This is also corroborated by  the fact that %kh
the charge current is strongly correlated with %kh
the spin current as a function of the phase difference between the two  %kh
superconducting banks.
We also investigate the asymmetric situation where
the thickness of one of the weak magnets is adjusted.
We again find that the equal-spin triplet pairs are responsible
for the flow of supercurrent and the spin current.
Next, we analyze the effect of misorientation angle between %kh
the exchange field in the half metal and the adjacent
weak magnet. 
When the misorientation  angle is $90^{\circ}$, %kh
the charge current, the equal-spin triplet pair amplitudes, 
and the spin currents attain their maximum values. 
On the other hand, when the angle is $0^{\circ}$ or $180^{\circ}$,
both the charge and spin currents vanish, showing
the importance of the magnetic configuration in half-metallic  %kh
Josephson junctions.
The induced triplet correlations also depend
on the exchange interaction for the central ferromagnet.
They saturate when the half-metallic limit is reached.
Finally, we show that when the exchange fields in the weak magnets
have the same magnitude and are perpendicular to that of the half metal,
the spin-valve effect is most pronounced.

\acknowledgments

C.-T.W. is supported by the MOST Grant No. 106-2112-M-009-001-MY2, the NCTS of Taiwan, R.O.C.
and a grant of HPC resources from NCHC.
K.H. is supported in part by ONR and a grant of HPC resources from the DOD HPCMP. 
K.H. would also like to thank M. Alidoust for useful discussions pertaining to this work. %kh2

\appendix

\section{Spin Rotations}
\label{appA}
Here we outline the spin rotations 
that are performed on 
the  triplet components ($f_0$, $f_1$, $f_2$)
in Eq.~(\ref{fall}).
By aligning the spin axes with the local exchange field directions,
the role of the triplet correlations and their physical interpretation becomes
clearer.
The central quantity that we use to perform the desired rotations
is the spin transformation matrix $\mathcal{T}$ in particle-hole
space. 
The quasiparticle amplitudes  transform as,
%${\bm \Psi} '' = T^{-1} {\bm \Psi}$):
\begin{align}
\Psi^\prime_n (x)
 = \mathcal{T}
 \Psi_n (x), \label{transform}
\end{align}
where $\Psi_n (x)=(u_{n\uparrow}(x),u_{n\downarrow}(x),v_{n\uparrow}(x), v_{n\downarrow} (x))$,
and the prime denotes quantities in the rotated system.
The matrix 
$\mathcal{T}$  can be written 
solely in terms of the angles that describe the local
magnetization orientation.
In particular, when the orientation
of the exchange fields in a given layer
is expressed in terms of the angles 
given in Eq.~(\ref{have}), we can write:
\begin{align}\label{tmatsmall}
\mathcal{T}= &\left[
  \begin{array}{cccc}
  \cos\left({\theta_i}/{2}\right)  &
    -i\sin({\theta_i}/{2}) & 0  &0\\
    -i\sin({\theta_i}/{2}) &
        \cos({\theta_i}/{2}) & 0 & 0\\
        0&0&\cos\left({\theta_i}/{2}\right)&
  -i\sin({\theta_i}/{2}) \\
  0&0& -i\sin({\theta_i}/{2})&
    \cos({\theta_i}/{2})
  \end{array}
\right].
\end{align}
Using the spin rotation matrix $\mathcal{T}$, it is also possible to transform the
original BdG equations ${\cal H}\Psi_n=\epsilon_n\Psi_n$ (Eq.~(\ref{bogo})) by performing the
unitary transformation: ${\cal H}' = \mathcal{T} {\cal H}
\mathcal{T}^{-1}$, with   $\mathcal{T}^\dagger \mathcal{T} =1$. 
As is the case under all unitary transformations,
the eigenvalues here
are preserved, but the eigenvectors are modified in general
according to Eq.~(\ref{transform}).
Thus we can write,
\begin{align}
u'_{n\uparrow}&=\cos\left({\theta_i}/{2}\right) u_{n\uparrow}-i\sin({\theta_i}/{2}) u_{n\downarrow}, \\
u'_{n\downarrow}&=\cos\left({\theta_i}/{2}\right) u_{n\downarrow}-i\sin({\theta_i}/{2}) u_{n\uparrow},\\
v'_{n\uparrow}&=\cos\left({\theta_i}/{2}\right) v_{n\uparrow}-i\sin({\theta_i}/{2}) v_{n\downarrow}, \\
v'_{n\downarrow}&=\cos\left({\theta_i}/{2}\right) v_{n\downarrow}-i\sin({\theta_i}/{2}) v_{n\uparrow}.
\end{align}

The terms involved in calculating the singlet pair
correlations (Eq.~(\ref{del})), thus
obey the following relation between the 
transformed (primed) 
and untransformed quantities: 
\begin{align}
{u'}_{n\uparrow} 
{v'}_{n\downarrow}^{*}+{u'}_{n\downarrow}{v'}_{n\uparrow}^{*} =
u_{n\uparrow} v_{n\downarrow}^* +u_{n\downarrow}
v_{n\uparrow}^*. 
\end{align}
Therefore  the terms that dictate the
singlet pairing are invariant
for any choice of quantization axis, transforming as scalars under
spin rotations.

The terms governing the triplet amplitudes on the other hand are in general 
 not invariant under spin-rotations. 
The relevant particle-hole products in Eq.~(\ref{f0})
that determine $f_0$, upon the spin transformations obey the following
relationships:
\begin{align}
{u'}_{n\uparrow} {v'}_{n\downarrow}^{*} -{u'}_{n\downarrow}
{v'}_{n\uparrow}^{*} &= \cos\theta_i (u_{n\uparrow} v_{n\downarrow}^{*}-u_{n\downarrow} v_{n\uparrow}^{*})\nonumber \\
&+i\sin\theta_i\left(u_{n\uparrow} v_{n\uparrow}^{*}-u_{n\downarrow} v_{n\downarrow}^{*}
\right), \nonumber \\
&=
f_0 \cos\theta_i 
+i\sin\theta_i f_2,
\end{align}
For the equal-spin component $f_1$ [Eq.~(\ref{f1})], the rotation leaves $f_1'$ unchanged:
\begin{align}
&{u'}_{n\uparrow} {v'}_{n\uparrow}^{*} + {u'}_{n\downarrow} {v'}_{n\downarrow}^{*}
=
%\cos\theta(u^{\uparrow}_{n} v^{\downarrow *}_{n}-u^{\downarrow}_{n} v^{\uparrow *}_{n})
%+\sin\theta(u^{\uparrow}_{n} v^{\uparrow *}_{n}+u^{\uparrow}_{n} v^{\uparrow *}_{n}).
u_{n\uparrow} v_{n\uparrow}^*
+u_{n\downarrow} v_{n\downarrow}^{*}.
\end{align}
For the other equal-spin component $f_2$  [Eq.~(\ref{f2})],   it is straightforward to show that
\begin{align}
{u'}_{n\uparrow} {v'}_{n\uparrow}^{*} -{u'}_{n\downarrow}
{v'}_{n\downarrow}^{*} 
&= \cos\theta_i (u_{n\uparrow} v_{n\uparrow}^{*}-u_{n\downarrow} v_{n\downarrow}^{*})\nonumber \\
&+i\sin\theta_i\left(u_{n\uparrow} v_{n\downarrow}^{*}-u_{n\downarrow} v_{n\uparrow}^{*}
\right),\nonumber \\
&=\cos\theta_i f_2
+i\sin\theta_i f_0.
\end{align}

\end{document}